\documentclass[review]{elsarticle}

\journal{Computer Communications}
\usepackage{graphicx}
\usepackage{booktabs}
\usepackage{float}
\usepackage{caption}
\usepackage{hyperref}
\usepackage{amssymb}
\usepackage{amsfonts}
\usepackage{amsthm}
\usepackage{booktabs}
\usepackage{amsmath}
\usepackage{balance}  
\usepackage{url}      
\usepackage[lofdepth,lotdepth]{subfig}

\renewcommand{\arraystretch}{1.2}
\newcommand{\ra}[1]{\renewcommand{\arraystretch}{#1}}

\begin{document}

\begin{frontmatter}
\title{\emph{CT-Mapper}: Mapping Sparse Multimodal Cellular Trajectories using a Multilayer Transportation Network}

\author[mainaddress]{Fereshteh Asgari}
\ead{fereshteh.asgari@telecom-sudparis.eu}
\author[mainaddress]{Alexis Sultan}
\ead{alexis.sultan@telecom-sudparis.eu}
\author[mainaddress]{Haoyi Xiong}
\ead{xhyccc@gmail.com}
\author[mainaddress]{Vincent Gauthier\corref{correspondingauthor}}
\ead{vincent.gauthier@telecom-sudparis}
\author[mainaddress]{Moun\^im A. El-Yacoubi}
\ead{mounim.el\_yacoubi@telecom-sudparis}

\cortext[correspondingauthor]{Corresponding author}

\address[mainaddress]{SAMOVAR, Telecom SudParis, CNRS, Université Paris-Saclay \\9 rue Charles Fourier 91011 Evry, France}

\begin{abstract}
Mobile phone data have recently become an attractive source of information about mobility behavior. Since cell phone data can be captured in a passive way for a large user population, they can be harnessed to collect well-sampled mobility information.
In this paper, we propose \textit{CT-Mapper}, an unsupervised algorithm that enables the mapping of mobile phone traces over a multimodal transport network. One of the main strengths of \textit{CT-Mapper} is its capability to map noisy sparse cellular multimodal trajectories over a multilayer transportation network where the layers have different physical properties and not only to map trajectories associated with a single layer. Such a network is modeled by a large multilayer graph in which the nodes correspond to metro/train stations or road intersections and edges correspond to connections between them.
The mapping problem is modeled by an unsupervised HMM where the observations correspond to sparse user mobile trajectories and the hidden states to the multilayer graph nodes. The HMM is unsupervised as the transition and emission probabilities are inferred using respectively the physical transportation properties and the information on the spatial coverage of antenna base stations.
To evaluate \textit{CT-Mapper} we collected cellular traces with their corresponding GPS trajectories for a group of volunteer users in Paris and vicinity (France). We show that \textit{CT-Mapper} is able to accurately retrieve the real cell phone user paths despite the sparsity of the observed trace trajectories. Furthermore  our transition probability model is up to $20\%$ more accurate than other naive models.
\end{abstract}


\end{frontmatter}

\section{Introduction}
\begin{figure*}
\centering
\subfloat[Road Trajectory]{
\includegraphics[width=0.48\textwidth]{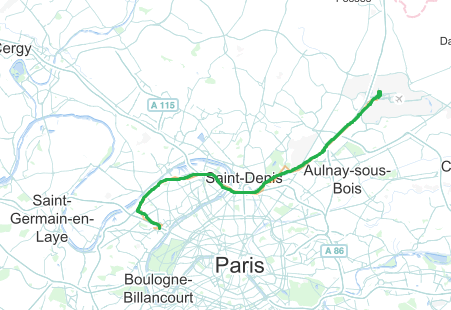}\label{fig:road}}
\subfloat[GPS Trajectory]{
\includegraphics[width=0.48\textwidth]{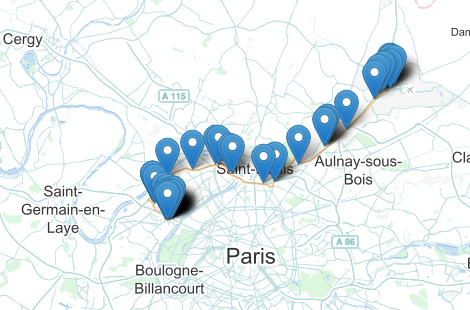}\label{fig:gps}}

\subfloat[Cell Trajectory (Full)]{
\includegraphics[width=0.48\textwidth]{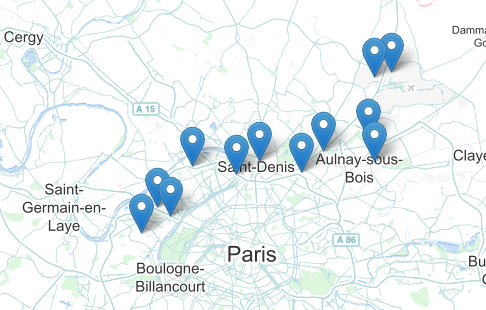}\label{fig:cell}}
\subfloat[CDR Trajectory]{
\includegraphics[width=0.48\textwidth]{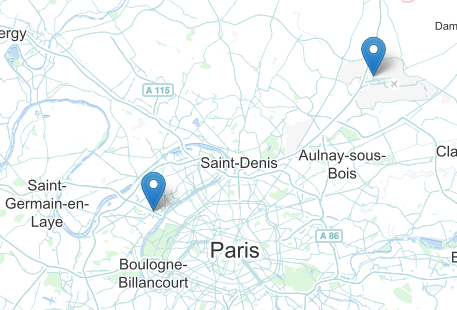}\label{fig:cdr}}

\subfloat[Sparse Cell Trajectory]{
\includegraphics[width=0.48\textwidth]{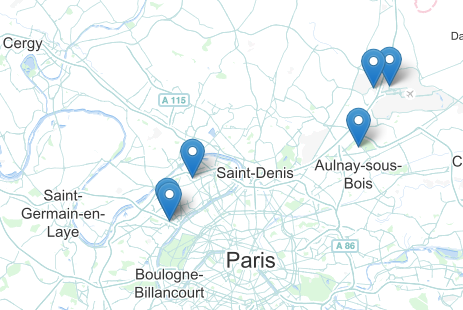}\label{fig:sparse}}

\caption{A user's trip from Airport CDG to city center of Paris: The road trajectory consists of the sequence of roads that the user passes-by; The GPS trajectory is sampled in minute based frequency; The Cellular trajectory (Full) records each cell tower the user passes-by; The CDR trajectory reports the location of the user's each call during the trip; The Sparse Cellular trajectory is sampled every 15 minutes}
\end{figure*}

Macroscopic analysis of the traffic flow in large metropolitan areas is a challenging task. This is especially true when multiple transit authorities are in charge of different transport networks (road, train, subway). Due to the lack of a common source of information across these transit systems, it is often hard for city authorities to grasp a unified view of mobility patterns. In this context, mobile phone data have recently become an attractive source of information about mobility behavior. Thanks to the ubiquitous usage of mobile phones, mining mobile phone data has become a promising way to understand  multimodal human mobility~\cite{reddy,Irish, actpas} ranging from identifying a mobile user daily path to recording transportation usage (e.g., taking train, metro, bus, etc.) in a large metropolitan area. Traditional approaches of mobility studies used GPS to accurately sense spatial data with a localization error bound $\leq$ 50m. Although it ensures the collection of fine-grained mobility trajectories (as shown in Fig.~\ref{fig:gps}), GPS-based data collection has two main drawbacks: first, it causes high energy consumption, and second, it is constrained to a limited  group of users (e.g. taxi drivers \cite{singapore} or a group of car drivers \cite{UnveilComplex}). GPS sensing, therefore, is not suitable for collecting large-scale data from metropolitan area populations. By contrast, cellular data provided by network operators does not suffer from these issues, and has become recently, as a result, a new source of mobility information. Signaling information from mobile network operators (CDRs -Call Data Records-) has been used as a valuable source of mobility information for large scale population \cite{actpas, d4d, Agarwal2015}.

Localization of mobile phone users with antennas (i.e., cellular towers), nonetheless, provides only coarse-grained mobility trajectories at antenna level, with a varying localization error of  hundred meters in densely populated cities, and within several kilometers in rural areas \cite{actpas}. Given the resulting \emph{cellular mobility trajectories} (i.e., a sequence of antenna \textit{id}s) and the location of each antenna as shown in Fig.~\ref{fig:cell}, it might be difficult to observe the road or metro station that the user passes by (as shown in Fig.~\ref{fig:road}).

In order to collect cellular mobility trajectories using mobile phones, previous works~\cite{d4d, actpas} usually extracted the trajectories from Call Detailed Records (CDR), where the CDR of a user restores the antenna \textit{id} and the time-stamp of each of his/her mobile calls. To understand human mobility, these works were mostly limited to aggregating the trajectories from a user's long-term CDR data  
in order to determine the frequently-visited locations and the visiting time (e.g., the park he/she usually passes by during the 07:00--09:00 window of working days). As such, the techniques proposed by previous works are not suitable for estimating the precise mobility trajectories on the road/transportation network using the CDR cellular trajectories.

Furthermore, one sample of CDR data (i.e., one call record) can be obtained only when the user places a call, making human mobility data between two consecutive calls irretrievable, especially when the time duration between the two calls is long (e.g., the inter-call mobility between the two calls in Fig.~\ref{fig:cdr}). Thus, even though it has been studied widely, CDR is unlikely to be a good data source for the trajectory mapping problem.
Considering the time sparsity drawbacks of CDRs,  we use, in this work, a new passive capturing technique to efficiently extract the position of the base stations the mobile phone is connected to. This technique analyzes the signaling channel of the data mobile network in order to extract the base station locations. This way of capturing the mobility of users is scalable and provides a higher sampling rate than CDR-based sensing.

The sparse cellular trajectories are collected and provided upon the request of the experiment participants to the network operator. Considering \emph{privacy issues}~\cite{privacy}, the network operator localizes each mobile user using an antenna \textit{id}, and further records each user's antenna \textit{id} with time-stamp periodically (e.g., every 15 minutes in our study). Compared to the user's real trip (in Fig.~\ref{fig:road}), the sparse cellular trajectory (in Fig.~\ref{fig:sparse}) partially measures the user's mobility with coarse-grained localization. The objective of our work is to map each sparse cellular trajectory~\footnote{In the rest of paper, we use the term "cellular trajectory" and "sparse cellular trajectory" interchangeably.} into the multimodal transportation network, in order to obtain the sequence of network nodes that the user passes by. For example, given the cellular trajectory shown in Fig.~\ref{fig:sparse} and the transportation network of the Paris metropolitan area shown in Fig.~\ref{fig:multilayer}, our goal is to recover the sequence of nodes of the real trip shown in Fig~\ref{fig:road}.

The common approach for mapping cellular trajectories into the metropolitan transportation (usually road) network is to first collect a large amount of cellular trajectories and then to manually label each cellular trajectory with the corresponding intersection sequence, an intersection being a graph node associated with a junction between two roads.
 The next phase is to train a \emph{supervised mobility model} (e.g., HMM) using the labeled cellular trajectories, in order to build a probabilistic model mapping antenna \textit{id} sequences to intersection sequences. After training, given a new user cellular trajectory, the supervised model predicts, as the mapping result, a sequence of intersections, having the maximal likelihood of generating the antenna \textit{id} sequence. However, labeling cellular trajectories to cover the road/transportation networks and all cellular towers of a metropolis is not practical, as it costs too much human efforts to collect the trajectories and to manually annotate them. We propose, in this paper, to solve the cellular trajectory mapping problem using an \textbf{ unsupervised mobility model}, that does not require collecting and labeling any trajectories.

Given the above examples and target research goals, the key issues in designing the unsupervised mobility model include:

1) \emph{Given the antenna id sequence in a cellular trajectory, retrieve the sequence of road/rail intersections that the user passes by given a \textbf{database} storing the \textbf{multimodal transportation network} - } The transportation network covering and connecting multiple types of transportation modes (e.g., rail, metro, highway , etc.) is named \emph{multimodal transportation network}~\cite{multimo}, in which each node is either a road intersection or a station of a rail transportation mode (i.e., subway, tramway and train), and each edge is a connection between intersections (e.g., the pathway connecting a metro station and a bus stop). Obviously, it is nontrivial to extract the precise user path from the multimodal transportation network using the antenna \textit{id} sequences.

The cellular trajectory might come from multiple transportation systems nearby each corresponding antenna and in different layers (underground, ground and trestle). To overcome this issue, it is necessary to build a comprehensive database storing all the intersections of the multimodal transportation network, where we can accurately retrieve the surrounding intersections of each antenna.

In this work, open data provided by OpenStreetMap (OSM) and the National Geographic Institute (IGN) are used to extract the multimodal transportation network of Ile-de-France (Paris and vicinity). This region is characterized by a high diversity of public transportation modes (tram, RER, train, bus) that have each particular specifications. Therefore, building a multimodal transportation network to study individuals' mobility requires a clear understanding of the multimodal network complexity. The multimodal transportation network is modeled in this work based on the concept of 'cross-layer' links that connect each two nodes where users can switch transportation modes.

2) \emph{Given an observed cellular trajectory, compute the \textbf{most-likely intersection sequence} over the multimodal transportation network  - } It is difficult to search the most-likely intersection sequence from the set of intersections, due to the following reason:

\textbf{Likelihood Computation:} In order to search the most-likely intersection sequence, given an observation sequence, we need to calculate the likelihood of each node given the cellular trajectory. While the traditional supervised HMM mobility model harnessing the statistics of labeled cellular trajectories (i.e. emission/transition probabilities) is usually used to estimate the likelihood, we propose an unsupervised HMM that does not leverage labeled data. Rather, it proposes a method to calculate the likelihood using the \emph{topological properties and other information of the transportation network}. In other words, the HMM parameters are automatically derived in an unsupervised way based on a priori knowledge of transportation network properties.

In summary, the main contributions of this work are:
\begin{itemize}
\item We propose to study the problem of mapping cellular trajectories to the \textit{multimodal} transportation network, in order to obtain the precise mobility of the users.  To the best of our knowledge, this is the first work addressing these issues. In particular, rather than mapping cellular trajectories using the supervised mapping algorithms with labeled mobility data, we propose an unsupervised mapping algorithm leveraging the topological properties of the transportation network, thus eliminating the tedious human labeling efforts for building the mobility model.

\item We propose an unsupervised trajectory mapping algorithm, namely \emph{CT-Mapper}, which maps cellular location data over the multimodal transportation network. The multimodal transportation network database was built using different references of geospatial resources.
The mapping algorithm is modeled by an HMM where the observations correspond to user cellular trajectories and the hidden states are associated with nodes of the multilayer graph. Transition probability and emission score were modeled based on topological properties of the transportation network and the spatial distribution of antenna base stations. The Viterbi decoding algorithm helps reduce the complexity of finding the best match which might enable us to deploy our unsupervised mapping algorithm on large scale mobility data sets in order to estimate multimodal traffic in metropolitan areas.

\item We collect real cellular trajectories of a group of users in the Paris metropolitan area with the help of a French telecom operator, then evaluate our mapping algorithm using the data. Through the extensive evaluation with  cellular trajectories covering more than  $2500$ intersection nodes and $3$ physical layers , $1000$ metro and subway stations, we show that our algorithm maps the cellular trajectory onto the multimodal transportation network of the Paris metropolitan area with good accuracy given the sparsity of user cellular trajectories.
This algorithm also achieves up to $20\%$ higher accuracy compared to a baseline approach, that exploits for unsupervised HMM parameter estimation, the complexity and topology of the multilayer network, without considering the transportation properties of network edges.

\end{itemize}

The rest of this paper is structured as follows: Sec.\ref{relatedwork} presents related work.
Sec. \ref{ov}  gives an overview of the proposed system. 
Sec. \ref{sec:transition} presents the details of the unsupervised estimation of HMM parameters and explains how the two main probability distributions used for mapping are derived. In Sec. \ref{evaluation} , we evaluate our proposed algorithm and the paper ends by a discussion and a conclusion in Sec. \ref{conclusion}.

\section{Related Work}\label{relatedwork}

\subsection{General Human Mobility Models}
A considerable amount of Human Mobility studies have been devoted to the analysis of trajectories of individuals based on their traces. Spatial characteristics such as the center of the mass, the radius of gyration and statistical characteristics revealed a number of scaling properties in human trajectories: Gonzalez et al \cite{barab} and Brockmann et al \cite{brock}  showed a truncated power-law tendency in the distribution of jump length.  It was observed that most individuals travel only over a short distance, and there is only a few who travel regularly over hundred kilometers. Further studies  \cite{uniModMob,barab} showed that travel patterns collapse into a single spatial probability distribution, indicating that, despite the diversity of their travel history, humans follow simple reproducible patterns. In addition, statistical analysis confirms that individuals' movement follows spatio-temporal patterns \cite{UnveilComplex,trajPatMin,PCA} which can help defining mobility models. In all mentioned studies, multimodal mobility aspects were not taken into account. One objective, in this work, is to investigate the mobility patterns of trajectories through the multimodal transportation network and to explore how these patterns are affected by the multiplicity of  the layers of the network.
Early mobility studies relied on expensive data collection methods, such as surveys and direct observation. Trajectories were mostly defined as Origin-Destination (OD), and were mapped over the desirable graph to retrieve an optimal path solution which is usually the shortest path between the Origin and Destination \cite{uniModMob,RoI,UnveilComplex,trajPatMin}.  
Although recent studies have tried to infer the traffic flow using additional traffic data \cite{traffic_flow_predict}, they still fail to retrieve the real path taken by individuals.    

\subsection{Mapping Algorithms}
Along with mobility studies, applications such as navigation systems, traffic monitoring and public transportation tracking, have used GPS data to track individuals or any moving object \cite{ctrack,Goh,hiden,kal,kalman, Hunter2011}. A variety of statistical approaches such as Expectation Maximization (EM) \cite{Hunter2011}, Kalman Filter \cite{kal,kalman} and Hidden Markov Model (HMM)  \cite{vtrack,ctrack,Goh,hiden,guidance} were used to map noisy sequential location data over transportation networks.  Most of these mapping algorithms have used GPS data as they provide accurate location data with an error of about 50 meters. 
Moreover, using labeled data, supervised models were trained to optimize model parameters in an automatic way. Once the models are trained, they are used to find the most likely path in the network assigned to sequences of noisy location data. Most of these mapping algorithms, however, were developed to map noisy data over road networks without considering other mobility modes.

\subsection{Human Mobility Modeling with CDR Cellular Trajectories}
Because of the expeditious growth of mobile phones, Call Data Records (CDR)  have been recently providing great data sets for human mobility studies as they are collected continuously for all active cellular phones. CDRs, however, have two significant limitations: first,  they are sparse in time because they are generated only when a phone engages in a voice call or text message exchange; and second, they are coarse in space and less precise than GPS location data, because they record location only at the granularity of a cellular antenna (with an average error of 175 meters in dense populated areas and up to 2 kilometers in non- dense areas). Nonetheless, the fact that almost the entire population is already equipped with cell phones \cite{actpas} allows for studying important aspects of individual mobility such as inferring transportation modes. Cellular network data were, for instance, used to classify different transportation modes for long-distance travels \cite{actpas,Irish}. Thiagaran et al. in \cite{ctrack} exploited cellular signal data with a combination of cellphone sensors to develop a supervised mapping algorithm in order to overcome the limitation of GPS data. While previous works have used cellular data to map long trajectories, this work proposes an unsupervised mapping algorithm that maps the sparse cellular trajectories over the multimodal transportation network in the Paris metropolitan area (Ile-de-France). This approach could be used for large scale 
smart-phone users for further studies in traffic estimation. Such a mapping is important for the development of smart cities and smart mobility. 

Studies of smart cities in the past were limited to analyzing multimodal transportation networks without considering large scale real mobility data.
The main goal of multimodal mobility studies is to improve public transportation monitoring and to reduce traffic congestion \cite{multimo, multimo1,electronic_ticketing}. Considering the aforementioned observations and the fact that the majority of trajectory mapping problems are developed for mono-modal transportation networks (specifically road networks), we believe that there is a gap in the literature. This study aims at bridging this gap by mapping cellular sparse data of smartphones over the multimodal transportation network in the Ile-de-France metropolitan area. The multimodal mapping results may help not only optimizing the multimodal transportation network, but also investigating the multimodal mobility behavior of individuals in metropolitan areas.

%
%
\section{CT-Mapper System Overview} \label{ov}

%
\begin{figure}
\centering
\includegraphics[width=0.9\textwidth]{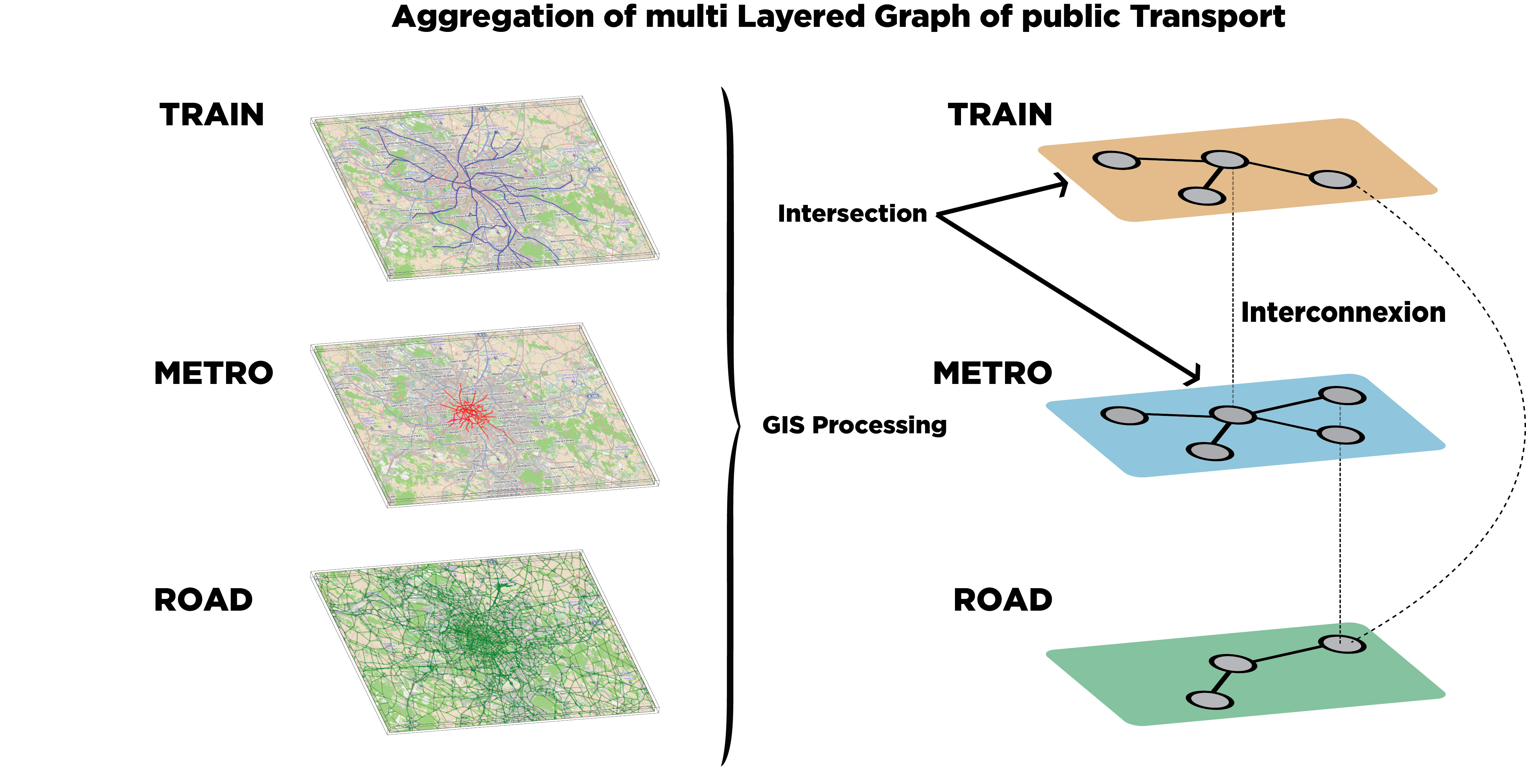}
\caption{ \small{Multilayer representation of different transportation networks} \label{fig:multilayer}}
\end{figure}

In this section, we first formulate the search problem of CT-Mapper, and introduce the dataset collected for mapping. We then analyze the computational complexity of the mapping problem over the collected dataset, and finally present the framework of CT-Mapper.

%
%
\subsection{Problem Statement}

In this section, we first formulate the problem by defining several key concepts used in our approach. 
 
\textbf{Definition 1. \emph{Multilayer Transportation Graph} -} Such a graph is represented as $\textbf{G}=(V,E,L,\Psi)$ where $V$, $E$ represent the vertices and the edges, $L$ is the set of possible layers.  
In our study we focused on 3 layers: road, train and subway.\\

\textit{Function $\Psi$} indicates the layer of each node $\Psi: V \rightarrow L$ in $\textbf{G}$. \\

\textit{Transportation Layer} $G^l=(V^ l,E ^ l)$ is a subset of $\textbf{G}$, where $V^{l}=\{v |v \in V, \Psi(v)=l\}$ and $E^{l}=\{<v_i,v_j> \in E, \Psi(v_i)=\Psi(v_j)=l\}$. Each node $v_i$ is characterized by its latitude and longitude (i.e., the geographical position $v_i=<lat,lon>_i$)\\

\textit{CrossLayer edge} set $E^{cl}\subset E$ defines the edges with pair of nodes not belonging to the same layer:  $E^{cl}=\{<v_i,v_j> \in \textbf{G} | \Psi(v_i) \neq \Psi(v_j)\}$\\

The multilayer Transportation graph is characterized by its \textit{adjacency matrix $W_{ij} \in \mathbb{R}^{|V| \times |V|}$}. 
Fig. \ref{fig:multilayer} illustrates how different transportation layers have been aggregated to build a multimodal transportation network. 

 \textbf{Definition 2. \emph{Cellular Network} - } In this work, we characterize a cellular network as a set of cell towers $C=\{c_0,c_1,...c_P\}$, where each cell tower $c_p=<lat,lon,r^{max}>_p$ is characterized by its latitude and longitude in the geographical coordinate system and by $r^{max}$ which is the maximum radius of the voronoi cell enclosing $c_p$ in the voronoi graph built from set $C$. Please note that the location of each cell tower does not coincide with the location of any intersection in the transportation network i.e., $\forall v^{}_i\in V$, $\forall c_p\in C$, we have $<lat,lon>_p  \neq <lat,lon>_i$.

%
\begin{figure}
\centering
\includegraphics[width=0.7\textwidth]{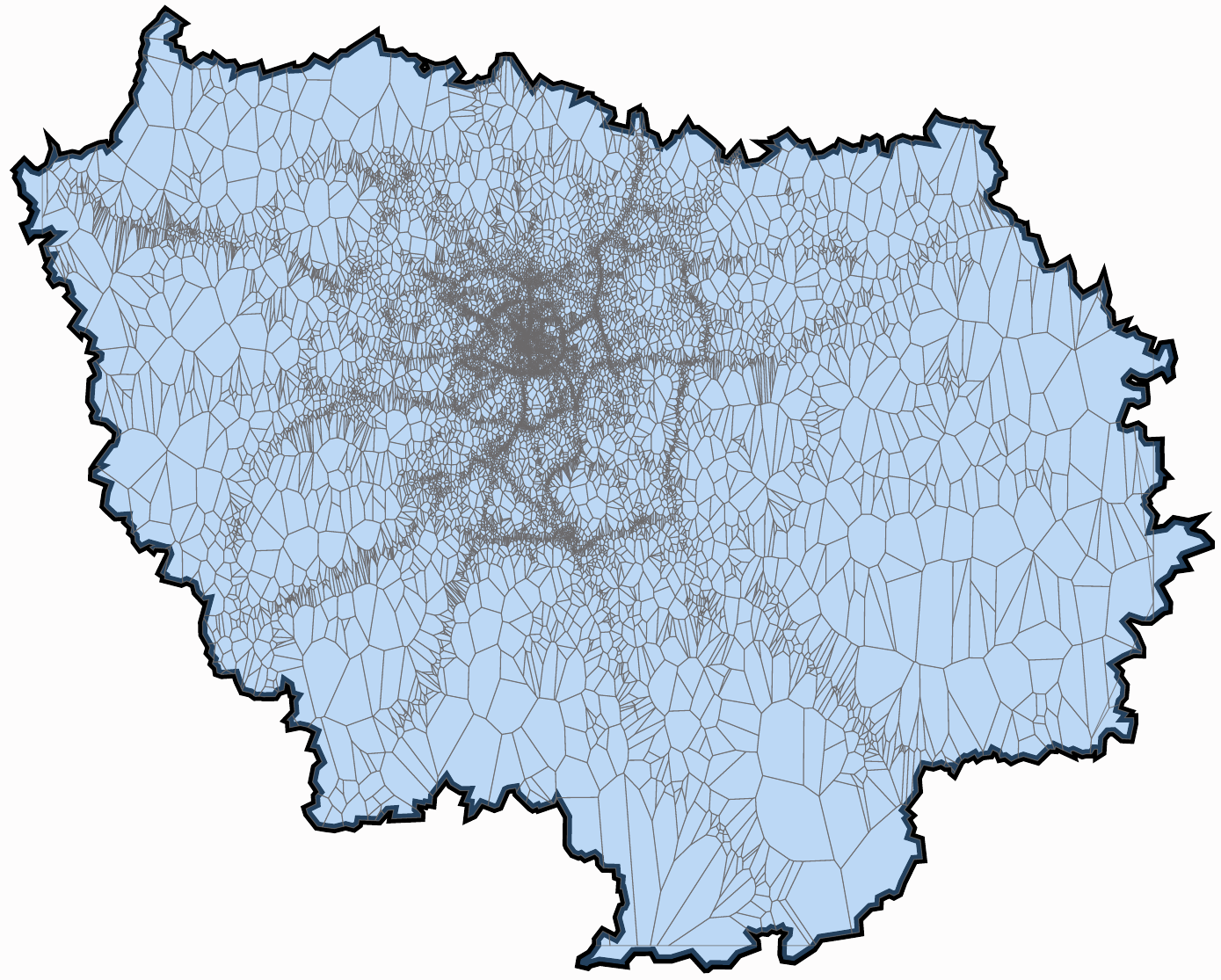}
\caption{Voronoi tessellation of cellular antennas in Ile-de-France\label{fig:voro}} 
\end{figure}

 \textbf{Definition 3. \emph{Sparse Cellular Trajectory} - }Further we define a {sparse cellular trajectory} of a user as a sequence of time-stamped locations $O=o_0\to o_1...\to o_M$, where each time-stamped location $o_t=<c(t)>$ refers to the cell tower at time-stamp $t$ the user is observed at.

 \textbf{Trajectory Mapping Problem -} Given a transportation network $\textbf{G}$, cell tower network $C$, and a user sparse cellular trajectory $O$, our search problem is to \emph{find a sequence of intersections $v_0 \to v_1...\to v_q$ which the user actually passes by on the transportation network}. 

%
%

\subsection{Data Collection and Datasets}
Three types of data are used in this study: multimodal transportation network data, sparse cellular trajectory data, and GPS trajectory data. The multimodal transportation network data are used to build the multilayer network graph and the mobility model for the mapping algorithms.  Cellular trajectories are used for testing while GPS trajectories are used as ground truth and not for training HMM parameters. 

\begin{figure}[htbp]
\centering
\includegraphics[width=0.8\textwidth]{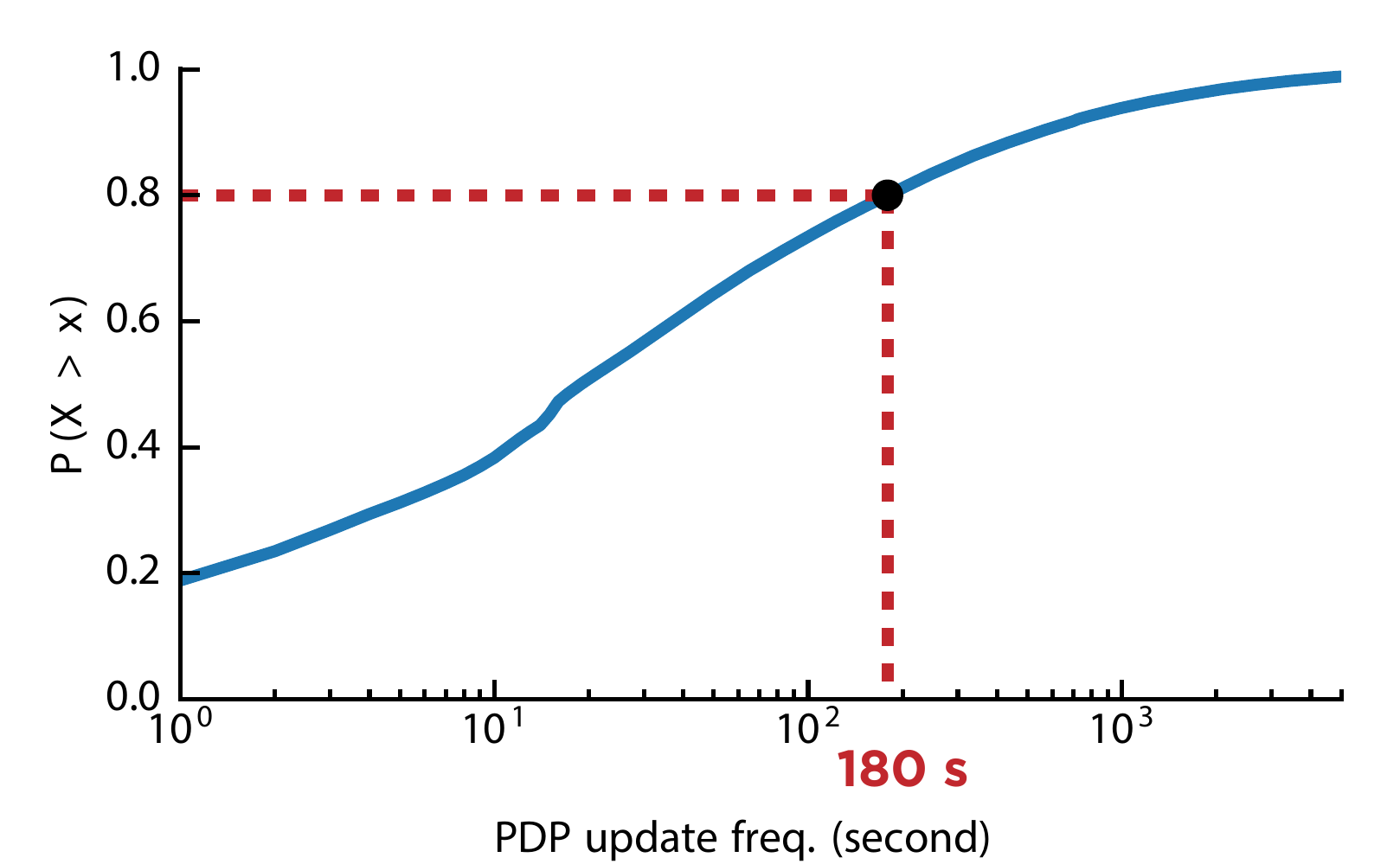}
\caption{PDP update frequency\label{fig:pdp_update_freq}}
\end{figure}

\textbf{Sparse Cellular Trajectory Data - } In this work we use a new type of cellular trajectory named Sparse Cellular Trajectory. A set of techniques for data collection are used to capture GPRS Tunneling Protocol (GTP) messages from the Cellular Data Network. Packet inspection of GTP-C (GTP control plane) enables us to capture users' localization information at higher frequency than the traditional CDR. The GTP is the tunneling protocol used to carry data traffic over the mobile network (from 2G to LTE) to internet.
When a smartphone enables its internet connection (e.g. when it is turned on), a message is sent over the network asking for access. This message contains among other things the identity of the phone and the cell id covering the user. Once the session is established, update messages are sent carrying information like the bearer or the cell id. These messages are triggered each time a change in the network parameters occurs: moving to another network area, change of bearer, and so on. More detail about the localization issue in GTP protocol can be found in \cite{Xu11}. However, unlike \cite{Xu11}, we found that the update occurs at a much higher frequency than described in \cite{Xu11} (cf. Fig. \ref{fig:pdp_update_freq}). Finally, when the mobile looses the signal or is turned off, a message closing the session is sent. With modern smartphone applications that emit and receive data on a regular basis (i.e. email, push notification), it is expected that the GTP tunnel for a given user remains constantly maintained, enabling us to sample the user position at each network event.

%

\textbf{GPS Trajectory Data - } To evaluate the accuracy of our proposed mapping algorithm, GPS data were used as ground truth. A group of participants were asked to install  the "Moves" smartphone application \cite{moves} to record their GPS locations. The GPS locations provided by "Moves" were analyzed to extract real trajectories of participants.

%
%

\subsection{Computational Complexity of the Mapping Problem in the Collected Datasets}
\begin{table*}[htbp]
\ra{1.3}
\centering
\begin{tabular}{@{}llllp{0.1\textwidth}l@{}}
 & \multicolumn{2}{c}{\textbf{Number}} & \multicolumn{2}{c}{\textbf{Avg.}} \\
\cmidrule(r){2-3} \cmidrule(l){4-5}  & {\small Node} & {\small Edge} &{\small  Degree} & {\small Length} &  {\small Reference} \\ 
\midrule
\textbf{Subway} & 303  & 356 & 2.35& 0.757 & OSM\\ 
\textbf{Train} & 241 & 244 & 2.025&3.07 & OSM    \\ 
\textbf{Road} & 14798& 22276& 3.01&1.34 &IGN\\
\bottomrule
\end{tabular}
\caption{ Different transportation networks with their properties}
\label{tab:multiplex}
\end {table*}

The underlying transportation network used in this study is the multimodal transportation network of Ile-de-France which is modeled by several separated graph layers corresponding each to a different transportation mode, interconnected together into a multilayer graph $\textbf{G}$. To build this graph, multiple geospatial datasets, namely the  road network from the National Geographic Institute (IGN)\cite{IGN} and the rail transport network (train and metro) from OpenStreetMap (OSM)\cite{osm} were aggregated. Each node in $\textbf{G}$ is either a road intersection, a rail station or a metro station. A key feature of the proposed multimodal transportation network is its modeling of transitions between different transport modes during a given trip. Cross-layer transition modeling is ensured by adding \textit{CrossLayer} appropriate edges between layers.  

Although such a multilayer representation of the transportation network enables us to model and define trajectories using different transportation modes, it also increases the complexity of the underlying network. To highlight this fact, we use the "search complexity " metric to show how difficult it is to find the sequence of segments that compose the truth path over the map. This metric describes how hard it is to find a sequence of nodes in a path from a source to a destination by chance. 

First, Table \ref{tab:multiplex} illustrates some topological differences between each layer in the multilayer graph $\textbf{G}$. For example, the average length between two consecutive intersections is rather heterogeneous across different transportation layers. To quantitatively assess the network complexity, we use an entropy measure to characterize the ease/difficulty of navigation in a network using "the search information" developed in \cite{0295-5075-69-5-853}, and in \cite{PhysRevLett.94.028701}. To summarize the work of \cite{0295-5075-69-5-853} and \cite{PhysRevLett.94.028701}, the search information entropy is the Shannon entropy of the probability of finding a given destination (in Eq. \ref{eq:path_entropy}) by chance; the higher is the entropy, the more difficult it will be for any search algorithm to find a right destination, regardless of its internal design.

Eq. (\ref{eq:path_entropy}) defines the probability for a random walker starting at node $s$ with degree $k_s$ to reach node $t$. Consequently, in Eq. (\ref{eq:average_entropy}), we define the search entropy of a graph as the sum over all shortest paths $\{SP_{st}\}$ from node $s$ to node $t$ in $G$ averaged over all possible pairs of nodes $(s,t)$ in graph $G$. As a result, by computing the average entropy of all the possible paths in $G$, we can express the relative complexity ($S_{avg}$) of finding a given path in a given graph $G$. 

\begin{figure}[htbp]
\centering
\includegraphics[width=0.8\textwidth]{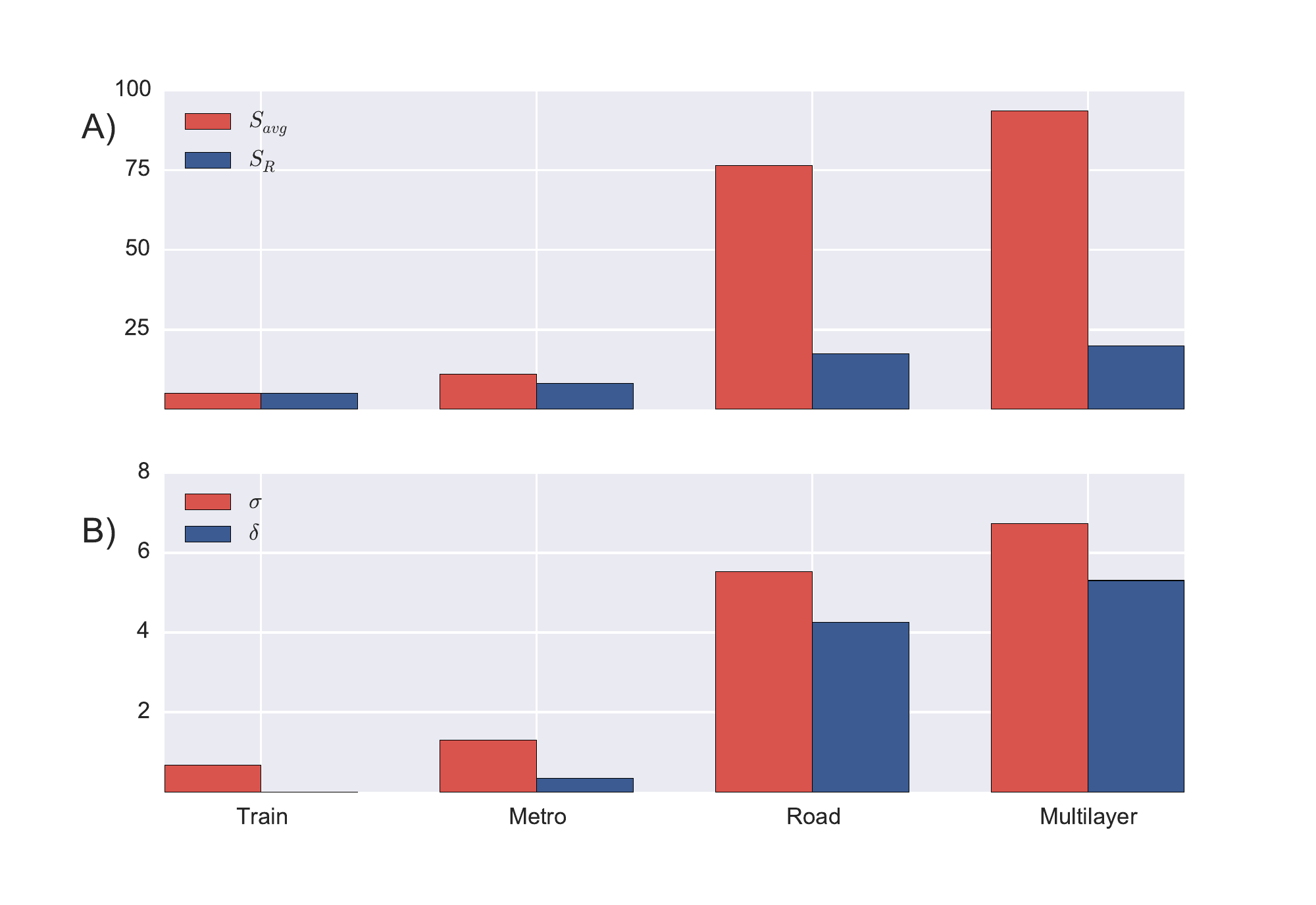}
\footnotesize{ \caption{Graph Entropy: (A) absolute value of the average entropy of the graph where $S_{avg}$ is the entropy of the real graph and $S_R$ is the entropy of the random graph with similar characteristics, (B) is the relative of the average graphs entropy of the paths in the subgraph of the metro, train, road} \label{fig:entropy}}
%
%
%
\end{figure}

\begin{equation}\label{eq:path_entropy}
P[SP_{st}] = \frac{1}{k_s} \prod_{j\in SP_{st}} \frac{1}{k_{j}-1}
\end{equation}

\begin{equation}\label{eq:average_entropy}
S_{avg} = \frac{1}{N(N-1)} \sum_{s = 1 }^{N} \sum_{t = 1 }^{N} -\log_2 \sum_{\{SP_{st}\}} P[SP_{st}]
\end{equation}

In Fig. \ref{fig:entropy}a, we plot the average entropy of each layer of the multimodal transportation graph of Ile-de-France along with the average entropy of the interconnected multilayer network. We observe that the average entropy is higher in the multilayer transportation network than in each of the layers taken separately. Fig. \ref{fig:entropy}b also shows the average path entropy relative to the size of the graph ($\sigma$). As it shows clearly, the complexity of the multilayer graph is higher than each of its layers taken separately, regardless of its size. We define $\sigma = S/\log_2(N)$ as the average graph path's entropy relative to its size and $\delta = (S_{avg} - S_R)/\log_2(N)$ to describe how a graph compares with its random counterpart in terms of its node degree, irrespective of the network size. 

As a conclusion, the search complexity of finding the right path in the multilayer transportation graph increases compared with a single layer graph. This is due to two effects: firstly when different layers are combined together in a multilayer graph, the number of degenerate paths (paths of the same length) increase and so does the overall complexity of the aggregate. Secondly, when we build the multilayer transportation network, we add multiple interconnections between each two layers, and we thus increase the degree of nodes that are at the junctions of two layers. It is also important to notice the clear increase of path complexity between the aggregate graph and the different layers taken separately (train, metro, road). The aggregation of layers increases the number of degenerated paths from typically one or two to several. 

These effects combined increase the search complexity of a given path in the multilayer transport network and increase, therefore, the difficulty of finding a correct mapping of the sparse trajectories on the graph. This phenomenon explains why in  multimodal transport systems using an algorithm that tries to find the best match  of a user trajectory (cellular trajectory) over the transport network will usually fails, due to the presence of many degenerate paths. 
%
%
%

%
%
\subsection{Framework and Overall Design}

\begin{figure*}
\centering
\subfloat[]{
\includegraphics*[width=0.19\textwidth,height=2.cm]{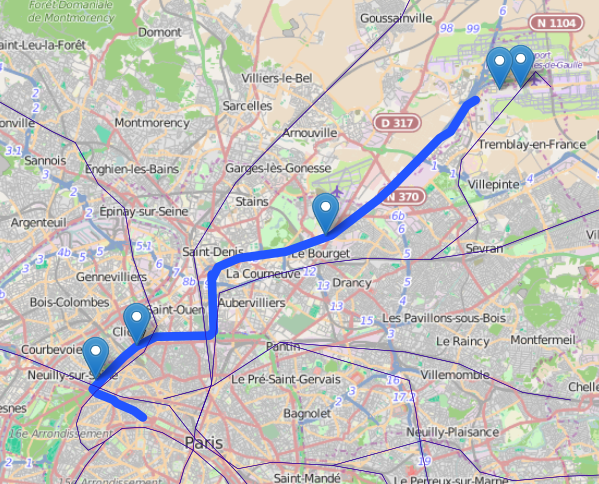}\label{fig:ph0}}
\subfloat[]{
\includegraphics[width=0.19\textwidth,height=2.cm]{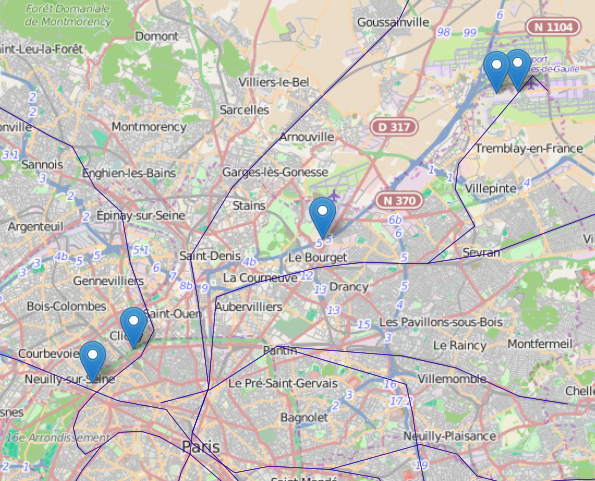}\label{fig:ph1in}}
\subfloat[]{
\includegraphics[width=0.195\textwidth,height=2.cm]{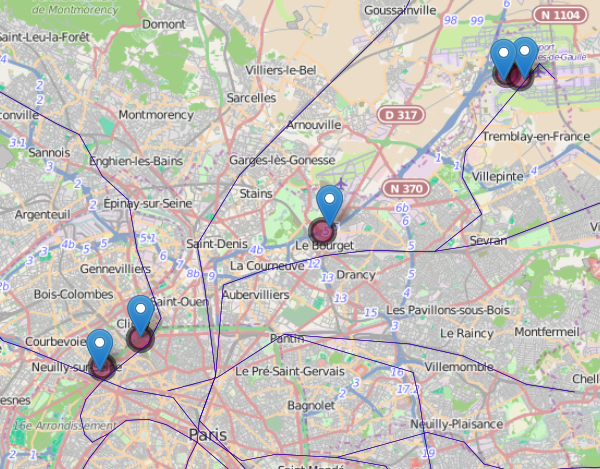}\label{fig:ph1out}}
\subfloat[]{
\includegraphics[width=0.19\textwidth,height=2.cm]{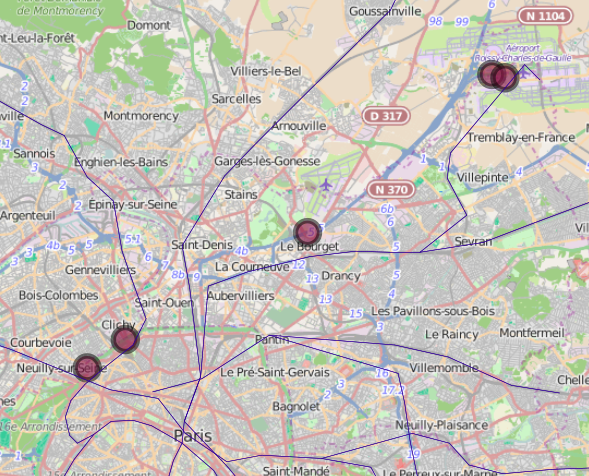}\label{fig:ph2in}}
\subfloat[]{
\includegraphics[width=0.19\textwidth,height=2.cm]{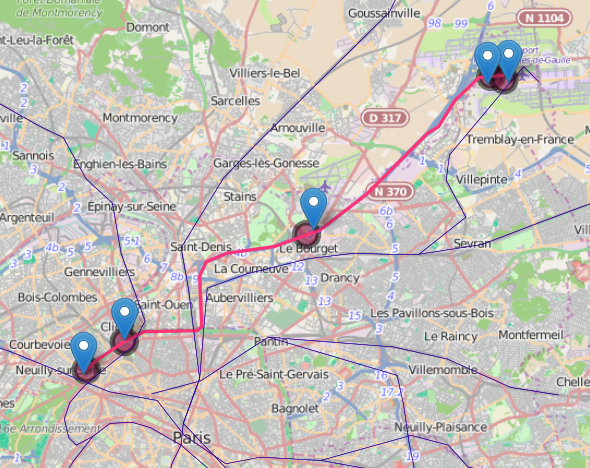}\label{fig:ph2out}}

\caption{An illustration of different phases of mapping algorithm a) Real Trajectory b) Cellular trajectory c) Phase I d) Phase II input d) Phase II output. The Blue line in the Fig. 5(a) is the real GPS trajectory of a user and given a sequence of 5 antenna base stations with the frequency of $15$ min, the mapping algorithm can retrieve the pink line in Fig.5(e)}
\end{figure*}

Given the multimodal transportation network $\textbf{G}$ and the cellular network $C$, we define an algorithm that outputs the most likely path or sequence of intersections given the sequence associated with a user sparse cellular trajectory $O$. In order to infer the accurate sequence of intersections from the given sparse cellular trajectory, we propose a \emph{two-phase} unsupervised mapping algorithm: in the \textbf{first phase}, the algorithm searches a sequence of intersections, namely the \emph{skeleton sequence}, where each two consecutive intersections are not necessarily adjacent (shown in Fig.~\ref{fig:ph1out}). For this objective, we developed an unsupervised Hidden Markov Model inference algorithm that accommodates the sparsity of observations (15 minutes).
The hidden states in the HMM are the multimodal graph nodes corresponding to road intersections or metro/train stations. The transition probability in our model takes care of sparsity of observations by permitting transitions between nonadjacent nodes as explained in Sec.\ref{sec:transition}.A . For each observation, a set of hidden states are selected as the candidate states in order to minimize the complexity of the search in the graph. Given a sequence of sparse cellular observations, our HMM model outputs the most likely sequence over the multilayer network. Note that for some sequences, only 3 or 4 observation points are available, which makes inference even more difficult. \\
Then, in the \textbf{second phase}, (shown in Fig.~\ref{fig:ph2in})  the algorithm traverses the {skeleton sequence} and outputs a sequence of adjacent intersections by completing the sequence (shown in Fig.~\ref{fig:ph2out}). Please note that the skeleton sequence searched in the first phase is with equal-length to the given sparse cellular trajectory $O$, while the intersection sequence outputted in the second phase would be longer than $O$. The reason is that, given the frequency of 15 minutes for observations, it is clear that a user would pass through more than one intersection between each 2 consecutive observation points, (e.g. when commuting with metro, it takes around 3 minutes to move between each 2 stations).

\textbf{Skeleton Sequence Search - } Given the sparse cellular trajectory $o_0\to o_1,...\to o_M$, this phase returns the skeleton sequence of the intersections as  $v_0\to v_1,...\to v_M$.
The algorithm is first initialized by $Pr_{t_0}(v_{i})=P(o_{0}|v_{i})$ for the candidate intersections  $v_{i}$ corresponding to the first time-stamped location $o_0$, with $Pr_{t_0}(v_{i})$ denoting the probability of a user to be located at intersection/node $(v_{i})$ at time $t_0$.
Then, for each candidate state corresponding to cell tower $o_t$,  the probability of a user being in $v_{j}$  at time $t$ and generating  $o_0 \to o_1 , ... \to o_t$  is calculated by Eq.~\ref{eq:greedy}; 
 \begin{equation}
Pr_t(v_{j})=  P(o_t|v_{j}) \times \max_{\forall~v_i} [Pr_{t-1}(v_{i}) \times Tr(v_{i},v_{j})] \label{eq:greedy}
\end{equation}

\noindent where $P(o_t|v_{j})$ is the  probability of a user connecting to cell tower of $o_t$ when he/she is in the intersection $v^{}_{j}$ and $Tr(v_i,v_j)$ is the transition probability of moving from node $v_i$ to node $v_j$. The parent node is also stored using Eq.~\ref{eq:parent}; 
 \begin{equation}
Par(v_j)=\displaystyle \mbox{arg}\max_{\forall~v_i } [Pr_{t-1}(v_{i}) \times Tr(v_{i},v_{j})] \label{eq:parent}
\end{equation}
At the end, we find 

$\overset{*}{v}_{M}= \displaystyle \mbox{arg}\max_{\forall v_M} Pr_t(v_M)$
Then a backtracking iteration using Eq.~\ref{eq:traceback}
 \begin{equation}
 \overset{*}{v}_{b-1}=Par(\overset{*}{v}_b) ~~~ \mbox{for}~~ b=[M,...,2,1] \label{eq:traceback}
\end{equation}
retrieves the most likely intersection sequence $\overset{*}{v}_0\to \overset{*}{v}_1,...\to \overset{*}{v}_{M}$ which produces the most likely path for the sparse cellular trajectory  $o_0\to o_1,...\to o_M$. 
Sequence  $ \overset{*}{v}_0\to \overset{*}{v}_1,...\to \overset{*}{v}_{M}$ serves as input for the next phase to retrieve the adjacent sequence of intersections for the given sparse cellular trajectory.

\textbf{Adjacent Sequence Completion - } Given the skeleton sequence $\overset{*}{v}_0\to \overset{*}{v}^{}_1...\to \overset{*}{v}_M$, for each pair of consecutive intersections $\overset{*}{v}_i,\overset{*}{v}_{i+1}$ that are not adjacent in multilayer graph $\textbf{G}$,
the algorithm searches the optimal sequence of intersections $v_{i_1}\to v_{i_2}...\to v_{i_k}$  and inserts the obtained optimal sub-sequence between the two intersections $\overset{*}{v}_i,\overset{*}{v}_{i+1}$ as:\\

\begin{equation}
 \overset{*}{v}_i\to  \underbrace{ 
v_{i_1}\to v_{i_2}...\to v_{i_k} }
\to \overset{*}{v}_{i+1} 
\end{equation} 

\[ \overset{\Uparrow}{ \mbox{{\footnotesize{Recovered path}}}} \]
as the complete adjacent sequence. Please note that each two consecutive nodes in the newly obtained sub-sequence  are adjacent in multilayer $\textbf{G}$. In the next section, we will introduce the calculation of probabilities used in our framework. 

\section{Core Algorithms} \label{sec:transition}
In the previous section, we described the general algorithm of mapping cellular trajectories over the multimodal transportation network. The two main probability distributions used in the mapping algorithm, are the HMM transition and emission scores that are estimated in an  unsupervised way. This section explains in detail how the two scores are defined and estimated.

\subsection{Transition Probability } 
The transition probability $Tr(v_i,v_j)$ in our mapping algorithm specifies the probability of an individual's moving from hidden state $v_i$ at time $t-1$ to hidden state $v_j$ at time $t$. 
The transition probability is inferred from the underlying network, the multilayer transportation network in which each transportation layer has its specific characteristics and properties.   
Table \ref{tab:multiplex} shows some graph topological properties such as the average node degree and the average physical edge length in different layers of the multimodal transportation network.
It is crucial to notice that relying on the topological properties of network layers without considering their differences, leads to a biased mapping algorithm in which the observations tend to be mapped over a specific transportation layer.
In addition, taking into account the sparseness of cellular observations, it is a key to authorize transitions between nonadjacent intersections.
We propose a transition probability of moving from intersection $v_i$ to intersection $v_j$ that is a function of 2 given factors:\\

1) Edge type and average speed over each edge: each physical edge in the multilayer graph $\textbf{G}$ belongs to a layer. Moreover, only the road layer contains different types of edges (such as highway, principal, local, etc.).      
We define matrix $W$  where each element of $W$ represents a weight between two nodes if there exists an interconnection between them. The weight of each link is defined as the inverse of average speed that one could have over the corresponding edge. Table \ref{tab:transition} shows the weight according to average speed over the edges of graph $\textbf{G}$.

 \begin{equation}
W_{ij}=\displaystyle \left\{
	\begin{array}{cl }
	 w_{ij} &~~~~ \mbox{if } v_i , v_j \mbox{ are adjacent in \textbf{G} }\\
     0 &~~~~ \mbox{otherwise.}                 
	\end{array} \right.   \label{weight}
\end{equation}

\begin{table}[htbp]
\ra{1.3}
\centering 
\begin{tabular}{@{}ll@{}}\toprule 
\textbf{$\mathbf{w_{ij}}$} & \textbf{Condition}  \\ 
\midrule
$1/80$ & $\Psi(v_i)= \Psi(v_j)=\mbox{metro}$ \\
$1/90$ & $\Psi(v_i)= \Psi(v_j)=\mbox{road (highway)}$ \\
$1/60$ & $\Psi(v_i)= \Psi(v_j)=\mbox{road (principale)}$ \\
$1/40$ & $\Psi(v_i)= \Psi(v_j)=\mbox{road (regional)}$ \\
$1/30$ & $\Psi(v_i)= \Psi(v_j)=\mbox{road (local)}$ \\
$1/10$ & $\Psi(v_i)\neq \Psi(v_j)=\mbox{crosslayer}$ \\
$1/100$ & $\Psi(v_i)= \Psi(v_j)=\mbox{train}$ \\
\bottomrule
\end{tabular}

\caption{\small{ Edge classification and weights for multilayer transportation network \textbf{G}.}}
\label{tab:transition}
\end {table}

2) Edge length: involving edge length in the transition probability indirectly considers higher probabilities for transitions to closer nodes. \\
The transition probability between two intersections $v_i$ and $v_j$ is defined as the inverse of the shortest path cost between $v_i$ and $v_j$:

\begin{equation}
 Tr(v_i,v_j)= \displaystyle \left( \sum_{\forall~(mn)~\in~SP{v_i v_j}}   w_{mn} \times d(v_m,v_n) \right) ^ {-1} \label{transition}
 \end{equation}

\noindent where $(mn)$ is the edge between $v_m$ and $v_n$ belonging to $SP_{v_iv_j}$, the shortest path between two nodes $v_i$ and $v_j$ in graph $\textbf{G}$. The shortest path cost of $SP_{v_iv_j}$ is the sum of distances over each edge $(mn)$ belonging to $SP_{v_iv_j}$, weighted by  $w_{mn}  $. $d(v_m,v_n)$ is the euclidean distance between each two nodes $v_m$ and $v_n$.  \\

In earlier studies, the transition probability was quantified based on topological properties of the underlying network which was mainly a road graph. 
In \cite{hiden,vtrack}, the transportation network was represented as road segments and transitions were assumed to occur between adjacent road segments. 
The authors in \cite{vtrack, guidance} considered equal transition probabilities between nodes in the same road segment or nodes between road segments which are adjacent with an intersection. The transition probability in \cite{ctrack} is defined based on the Manhattan distance between the grid cells of the road network. The objective of our proposed transition probability model is to minimize the bias of the mapping algorithm for layers with different topological properties.

\subsection{Emission Probability}

In HMM, at each time step $t$, there exists an observation $o_t$ which in our study is characterized as $c_t=<lon,lat,r_t^{max}>_t$. The emission score reflects the notion that it is more likely that a particular observation point is observed from a nearby intersection than from an intersection farther away \cite{vtrack}. For studies in which GPS data were used as observations \cite{vtrack,hiden,Goh}, the emission probability score is modeled by a normal distribution that is a function of the euclidean distance between the observation point and the hidden state, with a standard deviation estimated from sensor errors.

In this work, cellular antenna locations serve as observations; since there is no labeled data available to estimate cellular sensor errors, we build the Voronoi tessellation of cellular antennas in the area of study. 
In the Voronoi network of cellular antennas, each cellular antenna $C_i$ is characterized by radius $r_i$ which is the maximum distance of the cellular antenna from the corresponding Voronoi cell vertices. Our emission score is defined as a decreasing function of the distance between the antenna location and the hidden node (intersection):

\begin{equation}
Pr(o_t|v_j) \propto \displaystyle \left\{
	\begin{array}{cl }
     1.0 &~~~~~~ \mbox{if}:  d_{tj} \leq r_t^{max} \\
     \left( \displaystyle \dfrac{r_t^{max}}{d_{tj}} \right)^ \beta &~~~~~~  \mbox{if}:r_t^{max}  \leq d_{tj} \leq \tau \\
     0 &~~~~~~ \mbox{otherwise.}                 
	\end{array} \right.   \label{emission}
\end{equation}
                
\noindent where $d_{tj}=d(o_t,v_j)$ is the euclidean distance between $o_t$ and intersection $v_j$, and $\tau$ is a threshold corresponding to the maximum distance that a cell phone can be hit by a cellular antenna. $\tau$ enforces the constraint that only intersections in the radius of $\tau$ from the cellular antenna could be considered as candidate states (nodes).

\section{Evaluation} \label{evaluation}
\subsection{Dataset for Evaluation}

\begin{figure*}
\centering
\subfloat[Trajectory Time Distribution]{
\includegraphics[width=0.55\textwidth]{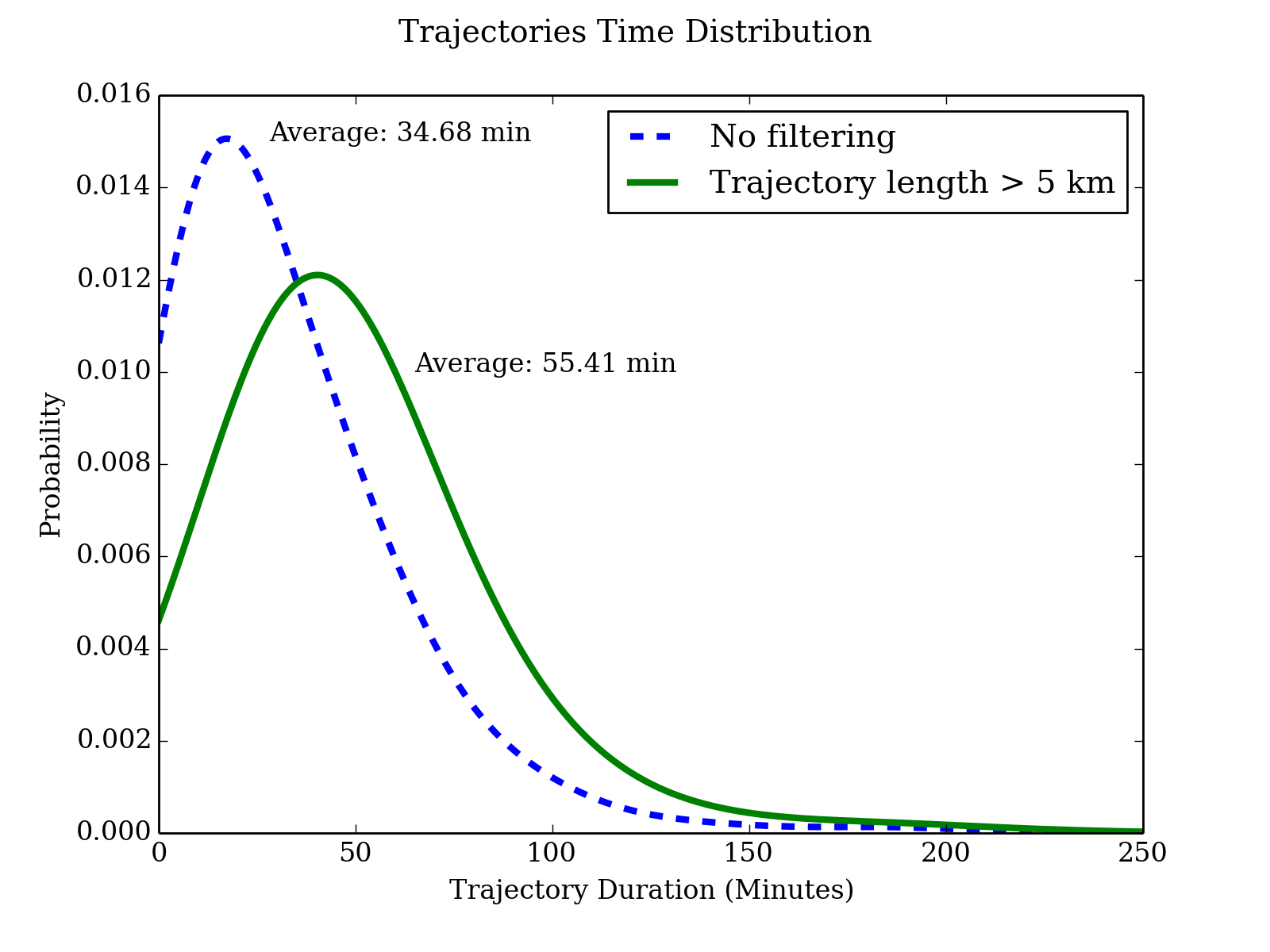}\label{fig:time_dist}}
\subfloat[Trajectory length Distribution]{
\includegraphics[width=0.55\textwidth]{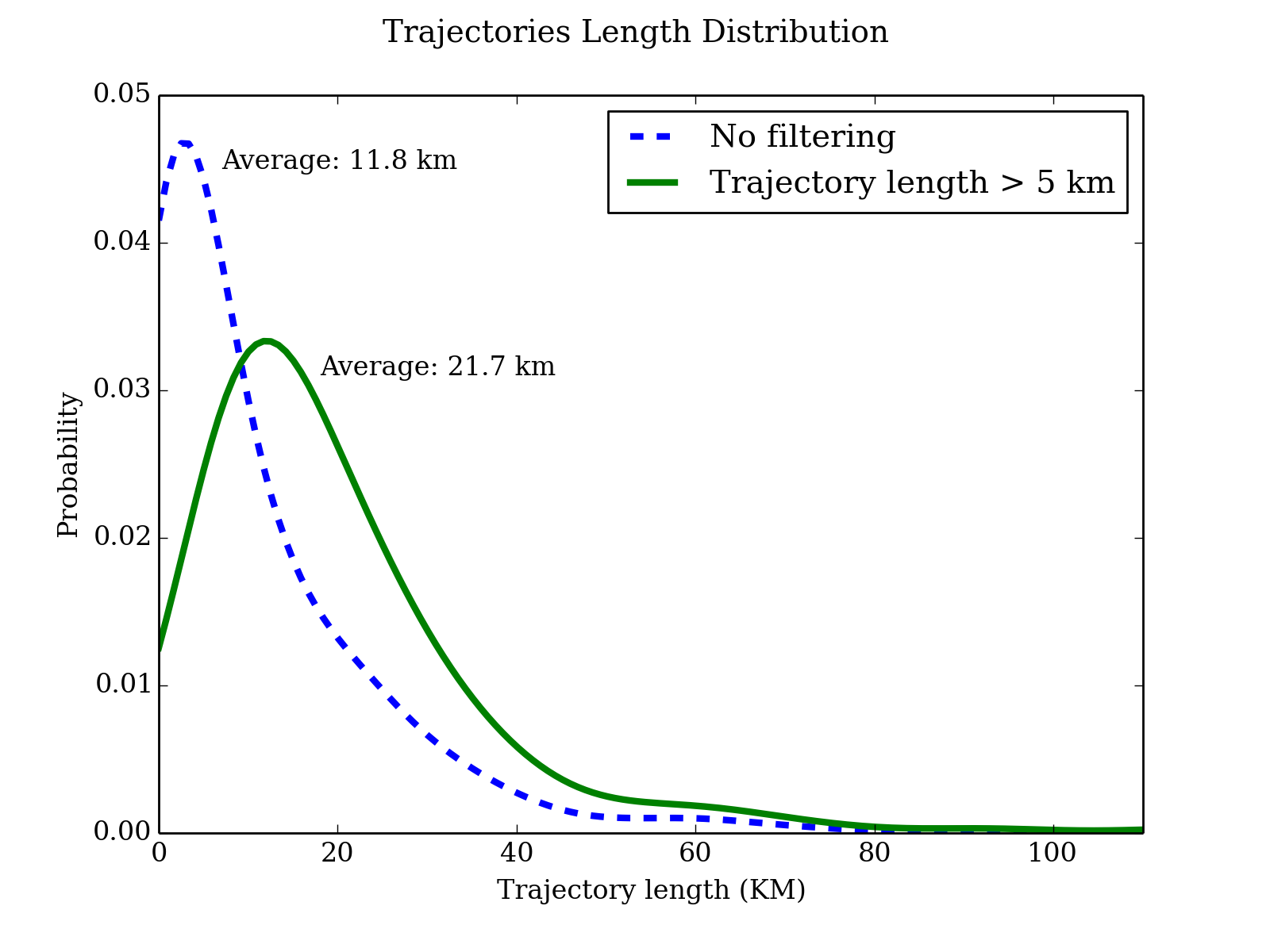}\label{fig:length_dist}}
\caption{ Time distribution and distance distribution}
\end{figure*}

\begin{figure}[htp]
\centering
\includegraphics[width=0.8\textwidth]{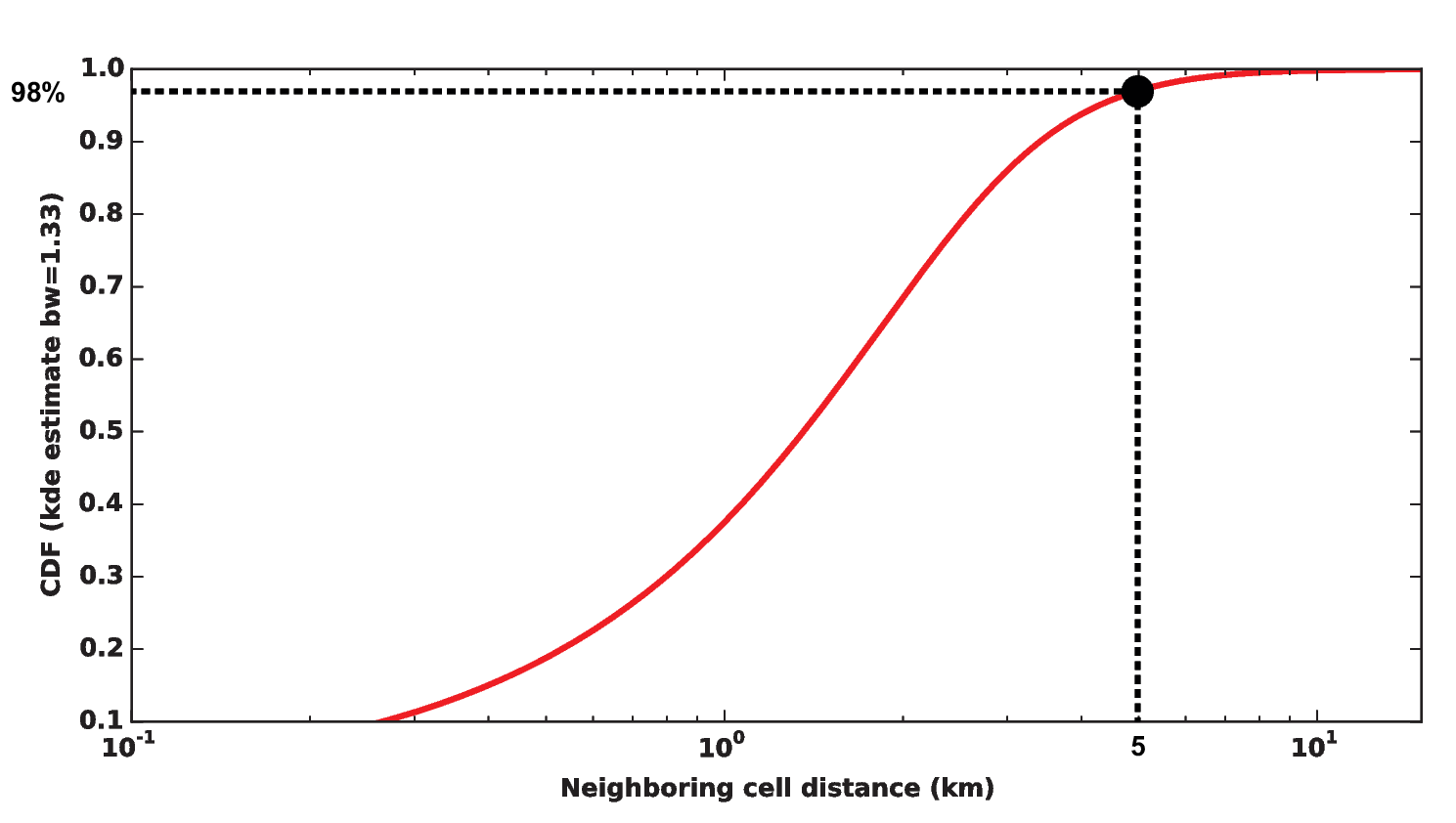}
\footnotesize{ \caption{Neighboring cell distance distribution} \label{fig:cdf_voro}}
\end{figure}

In order to evaluate the proposed algorithm, GPS data are used as ground truth.  We collected the cellular trajectories of 10 volunteer participants during one month (Aug-Sept 2014) with their corresponding GPS data. The GPS data were collected with the help of the application "Moves" \cite{moves}, which was installed on the participants' smartphones. The data captured were the sampled positions of the phone during its movements as well as its activities classified in four different categories: 'Walking', 'Running', 'Cycling' and 'Transport'. Based on this dataset, several prepossessing steps were performed in order to extract the trajectories to be mapped over the transport networks.

Trajectories whose lengths are shorter than $5$ kilometers were filtered out from the database. Given the low sampling rate of the cellular data (a data point every $15$ minutes), it is not realistic to seek recovering a movement with shorter duration than this threshold. The effect of this filter on the dataset distribution can be observed in Fig. \ref{fig:time_dist} and  Fig. \ref{fig:length_dist}. 

The spatial accuracy needed to distinguish a real mobility from noise depends on the distance between two base stations. In order to discard irrelevant movements, we filtered out all the trajectories under the threshold $x_{th}$ such that $P_{r}(X < x_{th}) = q $, where $P_{r}(X)$ is the distribution of distance between neighboring antennas. As Fig. \ref{fig:cdf_voro} shows, for $q= 0.97$, all the neighboring distances are less than $5$ kilometers.\\

\begin{figure}[htbp]
\centering
\includegraphics[width=0.7\textwidth]{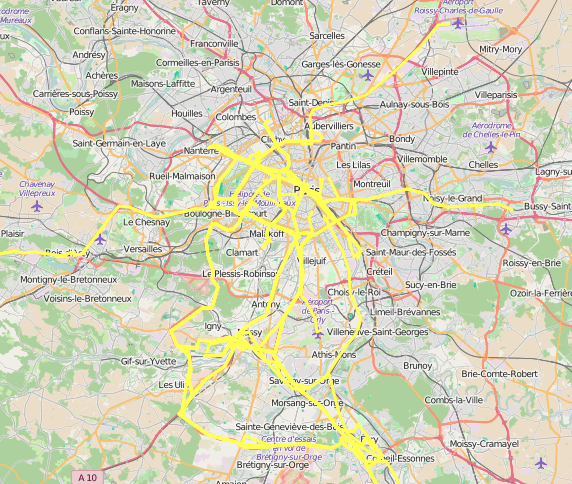}
\caption{ \small{The coverage area of GPS data collected is shown in yellow on the
map of Paris and region} \label{fig:cov}}
\end{figure}

To summarize, we built a dataset of 80 cellular trajectories (sequence of base stations) with their corresponding GPS paths mapped over a multilayer graph $\textbf{G}$. The multilayer transportation network contains around 16000 nodes and 26000 edges. The users' trajectories covered a total distance of $2200$ kilometers. The average number of observation points in each cellular trajectory is $5.55$ and the average length of a trajectory is $26.5$ kilometers. Fig. \ref{fig:cov} shows the coverage area of collected GPS trajectory dataset. 
%
%

\subsection{Evaluation Results and Comparison}

\begin{figure}[htbp]
\centering
\includegraphics[width=0.9\textwidth]{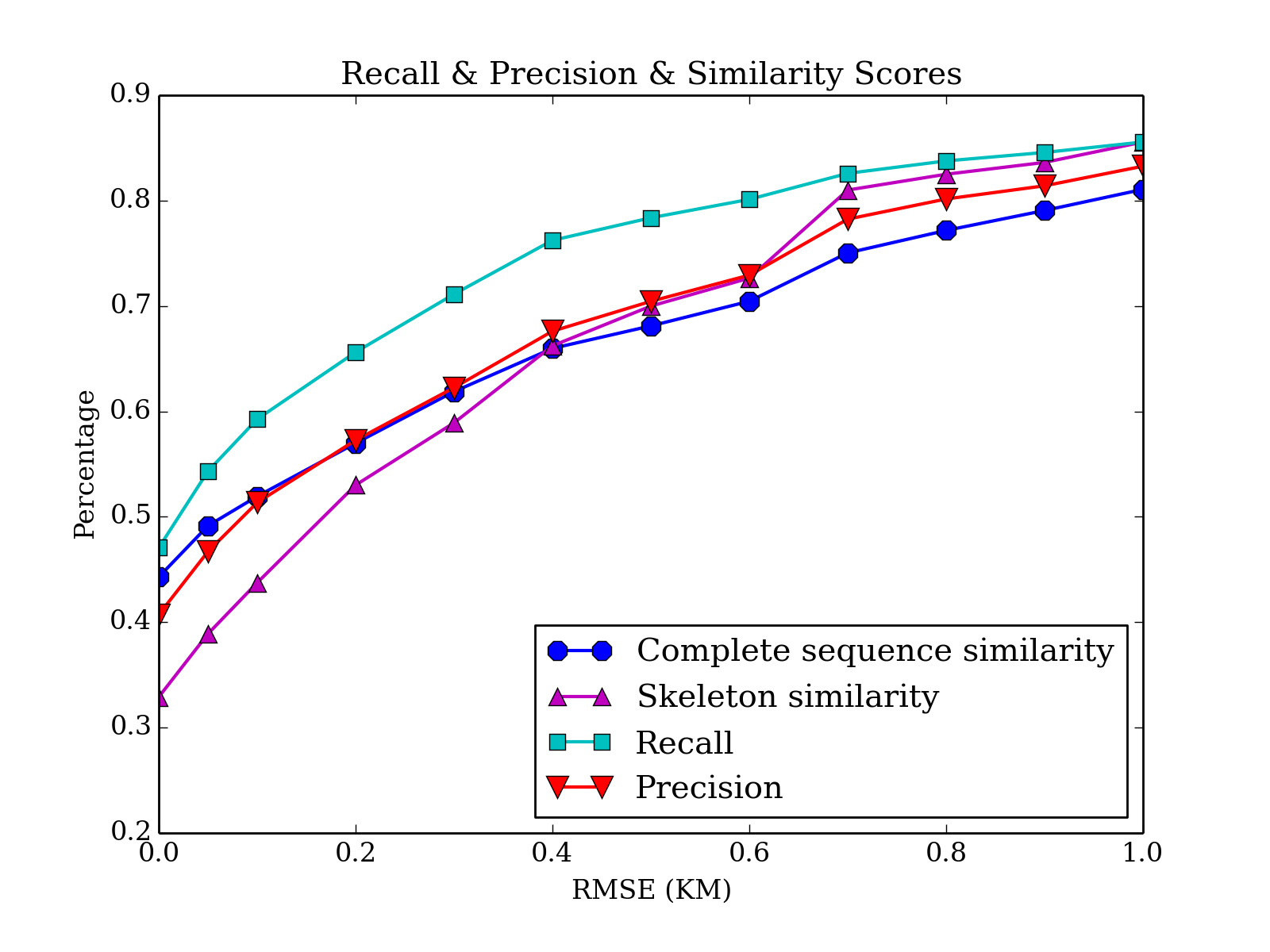}
\caption{ \small{Result evaluation} \label{fig:result1}}
\end{figure}

\subsubsection{Mapping algorithm efficiency}

To assess the effectiveness of our algorithm, the aforementioned labeled dataset was used for test and evaluation. We applied \textit{CT-Mapper} to map the cellular trajectories over the multimodal transportation network and to compare the result with GPS ground truth. 
Different measurements have been used to assess the performance of the Algorithm. First, we aim to quantify the similarity between the obtained path and the ground truth. Since the algorithm infers the real trajectory in two phases, the results of the mapping algorithm are evaluated in both phases accordingly. This similarity is quantified using the Edit distance score. This measure enables us to compare two sequences with different lengths by allowing different edits (deletion, insertion and substitution) to transform one sequence into the other. We evaluate the two phases of the algorithm by calculating the edit-based similarity scores for both the skeleton and the complete mapped sequence.
To have a comprehensive insight, we also calculate the average recall and precision of the results for dataset trajectories. Considering each trajectory as a set of nodes, precision is the fraction of retrieved nodes that belong to the real path. Recall (also known as sensitivity) is the fraction of correct nodes that are retrieved by the algorithm. 
Moreover, in the evaluation section, the Root Mean Square Error (RMSE) is used for two purposes: First, RMSE is computed to quantify the overall distance between the obtained result and the ground truth.  Second, owing to the considerable spatial noise of cellular observations, RMSE is used to detect matches between 2 points using threshold $\epsilon$ . In this case, if the RMS error between two points is smaller than $\epsilon$ , we consider the inferred point as a match.
For example, an error threshold of 0.1 kilometers indicates that for each node in the output sequence, the node is considered as a match point if it is within a 0.1 kilometer radius of its corresponding real location.
We calculated the four mentioned accuracy measures (precision, recall, skeleton and complete sequence similarity score) for a range of fixed allowed RMSE on the obtained mapping results.  The similarity scores are the complementary of the Edit distance scores. 
Fig.~\ref{fig:result1} shows the results of this evaluation.
For an allowed RMSE of 200 meters, we see that more than $50\%$ of skeleton and complete trajectories can be retrieved.
This is remarkable given the sparsity of the coarse grain cellular antenna positions with respect to user real trajectory (average of 5.5 observations per trajectory in the dataset while the average length is 26.5 $km$). It is important to mention that the frequency of cellular data collection is $15$ minutes and, therefore,  higher performances are expected if observations with higher frequencies are input to \textit{CT-Mapper}.
The average similarity score, for a RMSE of 1 kilometer, raises to $80\%$. In addition, \textit{CT-Mapper} reaches a recall and a precision of around $80\%$ when a RMSE of 1 kilometer is allowed. 
In addition to the metrics mentioned above, we compute the Edit distance error not as the number of required edits, but by considering the euclidean distance as the cost of each required edit. The average of Edit distances for all trajectories in the dataset is $0.79$ kilometer. \\

It is important to note that although the RMSE measures the overall spatial gap between the inferred path and the ground-truth path, the spatio-temporal information is implicitly taken into account. Assume, for instance, a road and a train route sections that are spatially similar, and assume an observation sequence that has similar length to and lies roughly between these road and train route sections. To a human observer, both an inferred road path or an inferred train path will look reasonable. However, as RMSE is the result of the comparison of the inferred path with the ground-truth path, the RMSE of the wrong inferred path would be much higher than that of the correctly inferred paths, especially if there are only few possible connections between these road and train route sections.

\subsubsection{Comparison with Baseline Algorithms}
\begin{figure*}
\centering
\includegraphics[width=1.0\textwidth]{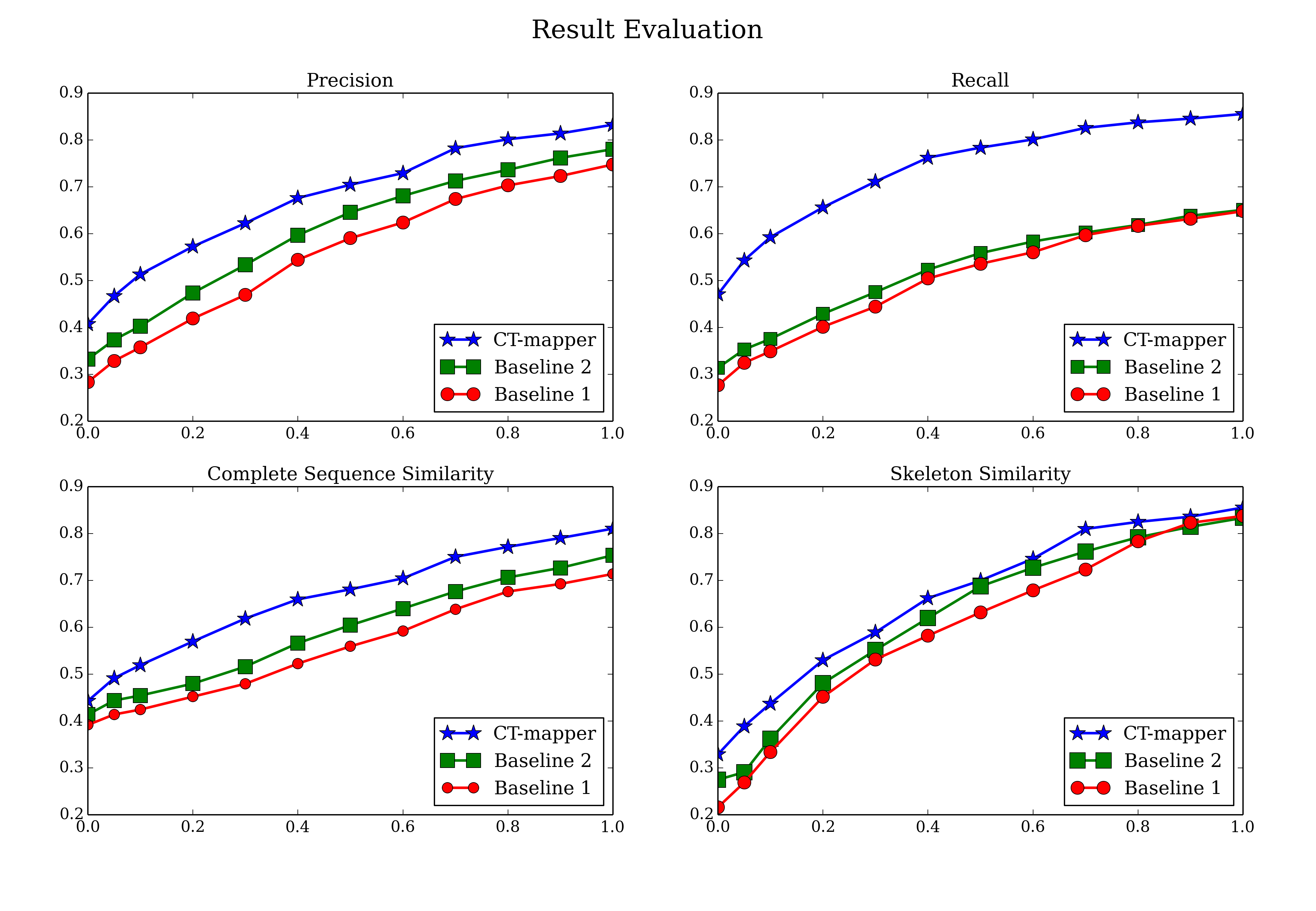}
 \caption {{Up-left: Precision, up-right: Recall, bottom-left is Edit-based similarity scores and bottom-left is the skeleton similarity score}  \label{fig:result}}
 \end{figure*}

In this section, the performance of our proposed model is compared with two baseline models. \textbf{Baseline 1} is a simple model that snaps each observation to the nearest node in the network to find the skeleton and for the second phase, uses least-cost paths between them to retrieve the full path. The result of this baseline model is compared with \textit{CT-Mapper} in Fig.\ref{fig:result}. \\
To evaluate our transition probability model based on transportation properties as presented in Eq.(\ref{transition}), we derive \textbf{Baseline 2}, an HMM based baseline model associated with the naive assumption consisting of setting equal probabilities for all outgoing transitions from each node (including self node transition). Under such a model, the transition probability between two nodes $v_i$ and $v_j$ is represented as:
\begin{equation}
Tr(v_i,v_{j})=\displaystyle  \left( {k_i} *  \prod_{n \in Q} k_{ n} \right) ^ {-1} 
\end{equation}

where  $Q=SP_{v_i v_j}-\{v_i,v_j\}$ and $k_i$ is the degree of $v_i$. 
This naive assumption considers all the multilayer network edges on equal footing irrespective of their layer transportation properties . 

\begin{figure}[htbp]
\centering
\includegraphics[width=0.9\textwidth]{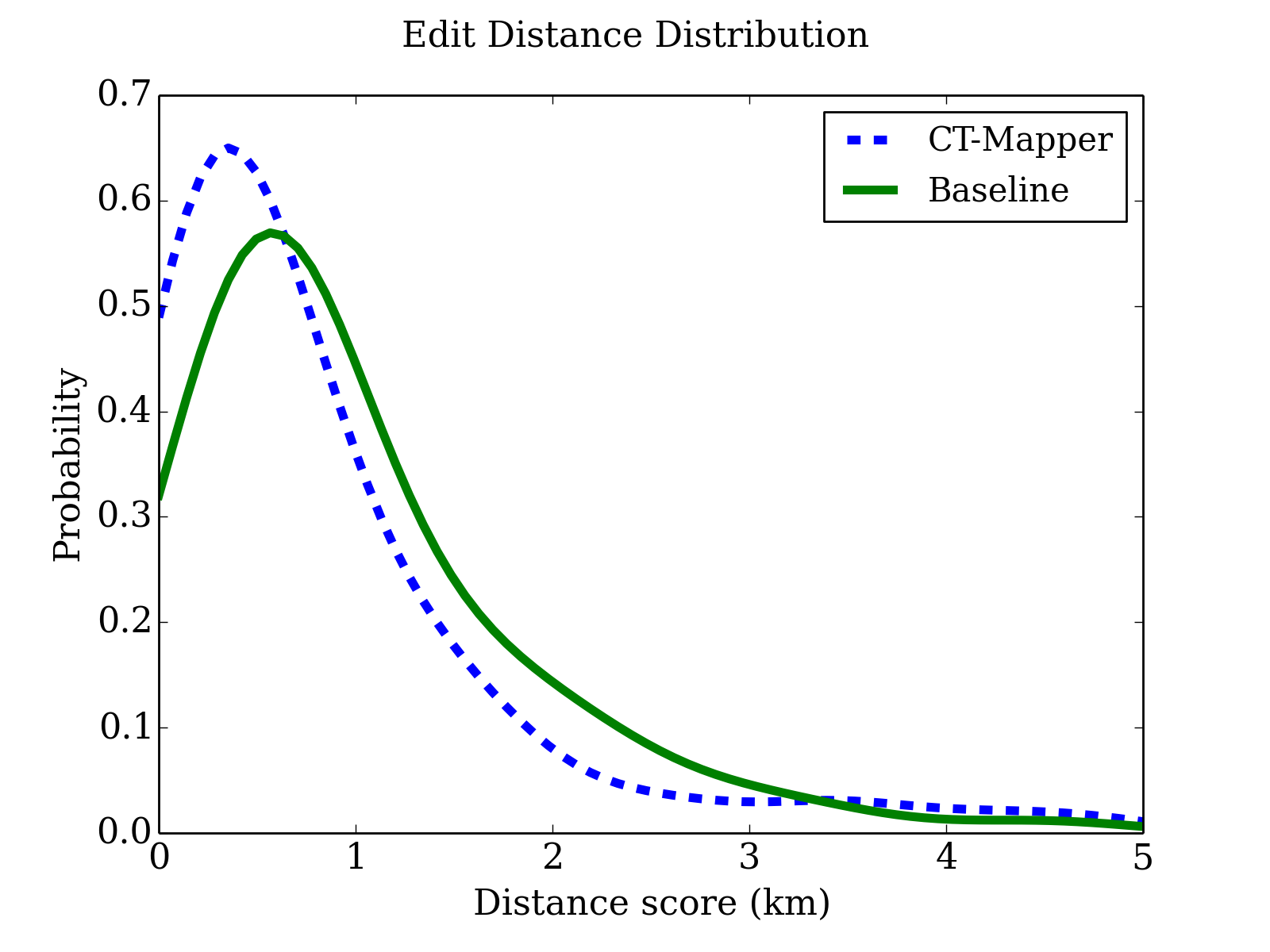}
\footnotesize{ \caption{Sequence Edist Distance} \label{fig:edit}}
\end{figure}

Using this transition probability model, we build an HMM in the same way as \textit{CT-Mapper} was developed. We use this model as a baseline algorithm and run it on the test dataset to compare the results with \textit{CT-Mapper}. 
We calculate the four performance measures for the baseline models.  Fig.~\ref{fig:result} compares the performances of the two models with \textit{CT-Mapper}. As the figures show, there is up to $20\%$ improvement in recall using our proposed transition probability model. 
Also, the average Edit distance of the baseline algorithm result was $1.04$ kilometer, which proves that \textit{CT-Mapper} performs significantly better compared to the second baseline algorithm. Fig.~\ref{fig:edit} shows the distribution of Edit Distance for both the second baseline algorithm and \textit{CT-Mapper}. 

\subsubsection{Multimodality analysis}

\begin{figure}[htbp]
\centering
\includegraphics[width=0.9\textwidth]{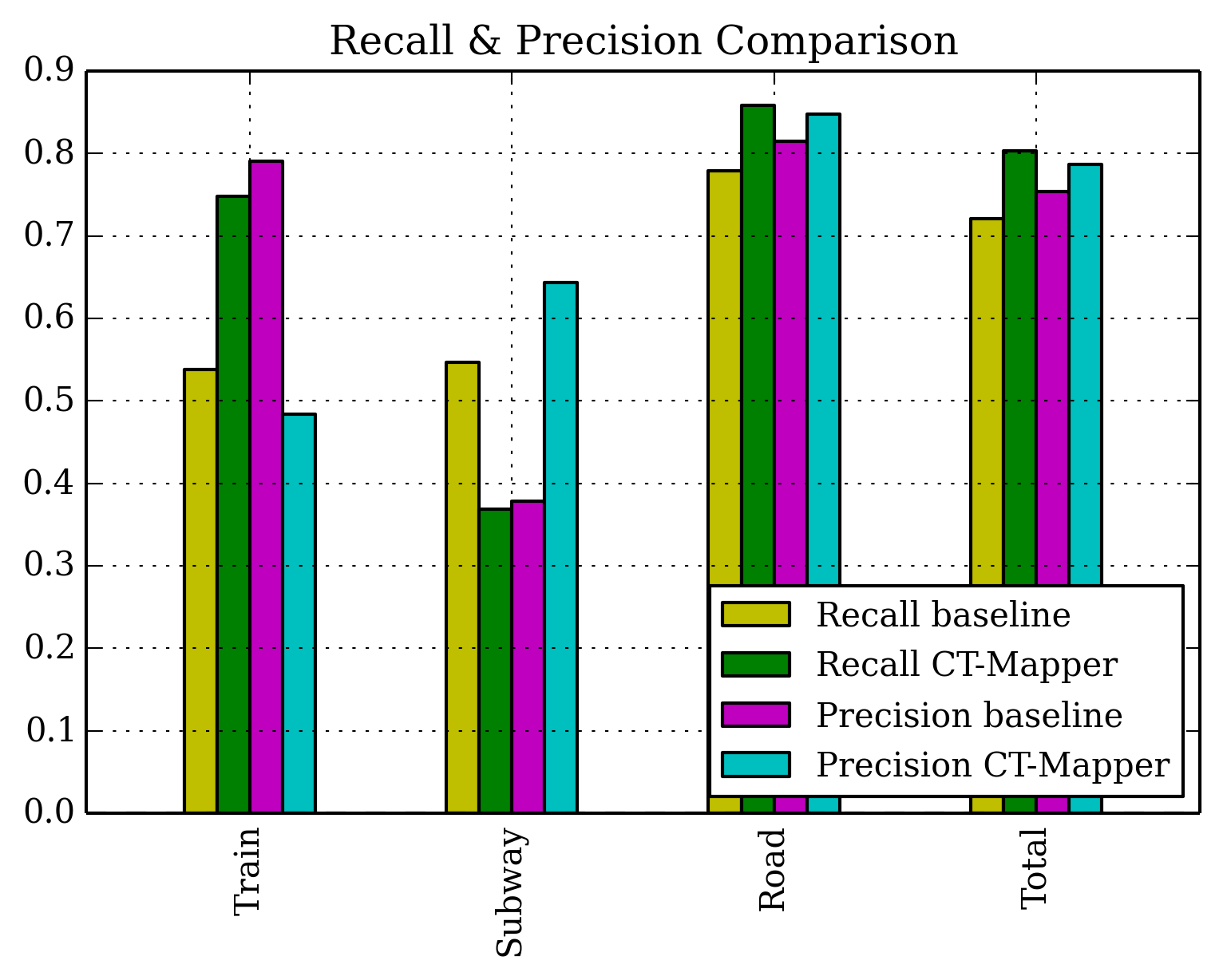}
\footnotesize{ \caption{Recall and precision in layer detection} \label{fig:barplot}}
\end{figure}
 
In the next step  of assessing our mapping algorithm, we investigate the accuracy of the mapping algorithm in transportation layer detection.  As mentioned in Sec.  \ref{transition}, the complexity of multimodal mapping significantly increases owing to the considerable topological differences between transportation layers. This issue is dealt with in the proposed transition probability model that seeks minimizing the bias in the mapping algorithm.

We calculate the recall and precision for correct layer detection  for each layer. The overall recall and precision for the whole network is computed as the average of recall and precision for each layer, weighted by the number of nodes.

As seen in Fig.~\ref{fig:barplot}, in some few cases (such as in the subway) the baseline algorithm performs better, but overall the \textit{CT-Mapper} algorithm is significantly better in terms of recall and precision.  

%
%
\section{Discussion \& Conclusion} \label{conclusion}

In this study, we proposed an unsupervised mapping algorithm \textit{(CT-Mapper)} to map sparse cellular trajectories over a multimodal transportation network. We modeled and built the multilayer transportation network of subway, train and road layers for the Ile-de-France (Paris) metropolitan area. The multilayer transportation network contains around $16000$ nodes and $26000$ edges. Investigating the complexity of the multilayer transportation graph, a transition probability model leveraging the transportation layer type and topological properties was estimated and used in an unsupervised HMM-based mapping algorithm. We carried experiments on a test dataset of $80$ real multimodal trajectories collected from $10$ participants during one month (Aug-Sept 2014) to evaluate our algorithm. Considering the sparsity of cellular observations (with a frequency of 15 minutes), the percentage of retrieved paths of smartphone users is notable. To validate our transition probability model, we compared it with a baseline algorithm that does not take into account the transportation properties of each layer. The results show up to $20\%$ of accuracy improvement of the first over the second. This shows that our model better accommodates the complexity of the multimodal transportation network.

Our model considers a transition probability between nodes that is inversely proportional to the average expected time over the paths between these two nodes. Nonetheless, our model does not take into account the time information stemming from the input observation sequence associated with a user trajectory. One of the reasons is that we use Sparse Cellular trajectories of data points sampled every 15 minutes. In this data sampling setting, the Telecom operator provides us only with the information on whether a data point is observed or not, not with its precise temporal occurrence within the considered 15 min time stamp. This makes the time information unreliable. That said, for new datasets with an accurate temporal information on the sampled data points,  we could harness the time information, by using, for instance, a Conditional Random Field (\textit{CRF}) to explicitly model the transition probability to a node at time \textit{t} conditionally on a node at time \textit{t-}1, and the previous observations, unlike the \textit{HMM} where such a probability is conditional only on the node at time \textit{t}-1. To do this however, we need a larger dataset as \textit{CRF} is more prone to overfitting.

Along the same lines, We expect that using a dynamic weight matrix, which is compatible with the traffic model at different times of the day, is likely to enhance the mapping results. This issue will be investigated in future studies. Furthermore, The improvement of accuracy measures of our mapping algorithm by minimizing bias mainly emanating from the multimodality of the transportation network is of great importance which shall be discussed in future contributions.  Finally, investigating the possibility of using the proposed mapping algorithm at near real-time (NRT) for traffic monitoring is another direction of further contributions.  

\section*{Ethics requirements and legal requirements followed during the data collection}
Before starting the experiments of collecting cellular data, we submitted the experiment protocol to the university ethics committee. Once the experiment started, each volunteer signed a legal agreement stipulating that each of them requested access to their cellular localization data for one month (with a sampling interval of 15 min). This request was bond with a legal agreement giving us the right to use their data for research purpose only. After one month of retention period after the end of the experiment, the cellular data were directly provided to the volunteers by the telecom operator. They forwarded us their cellular data afterward as well as permission to access their GPS traces.

\section*{Acknowledgments}
The authors would like to thank the reviewers for their useful comment and shepherding the paper, they have certainly helped us to enhance this paper. We would also thanks Marco Fiore for his helpful discussion on this topic. This research was sponsored in part by the Pierre and Marie Curie University and Telecom SudParis through their PhD founding program.

\section*{ References}
\bibliography{bib}

\appendix

\section{Example of traces where our map-matching partially fail to retrieve to true path}
\begin{figure*}
\centering
\subfloat[Mapping failure 1]{
\includegraphics[width=0.49\textwidth]{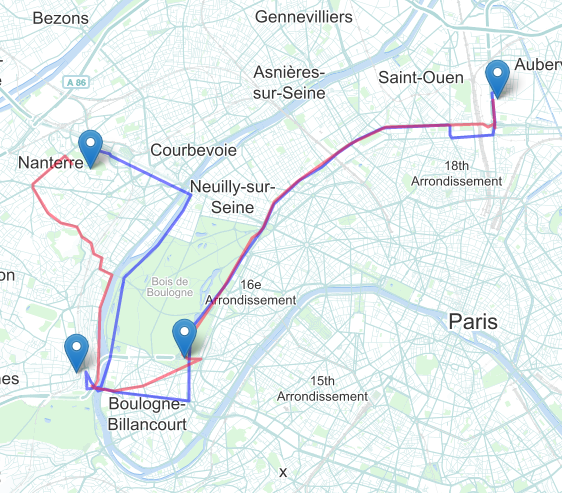}\label{fig:mf0}}
\subfloat[A correct mapping result]{
\includegraphics[width=0.49\textwidth]{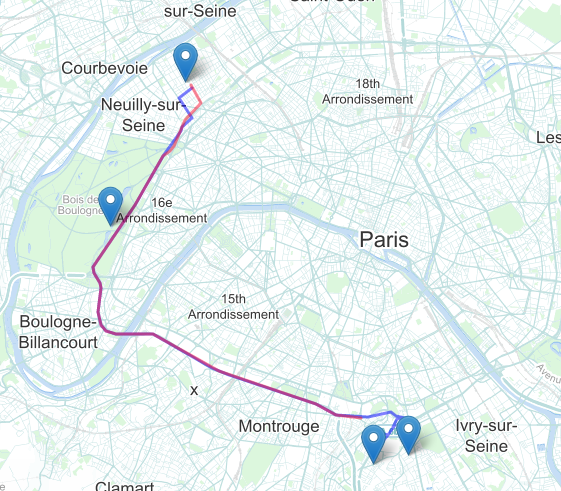}\label{fig:mf2}}

\subfloat[Mapping failure 2]{
\includegraphics[width=0.49\textwidth]{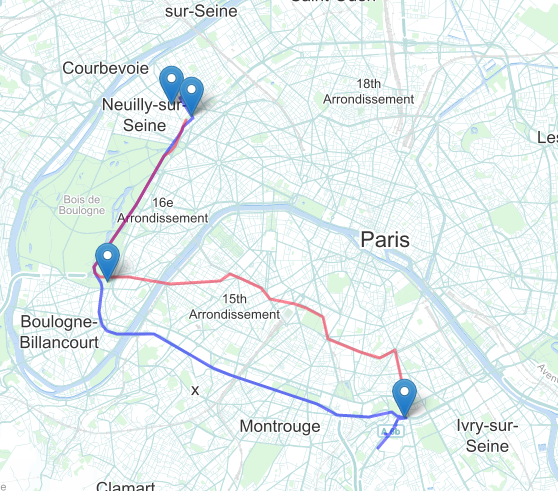}\label{fig:mf3}}
\subfloat[Mapping failure 3]{
\includegraphics[width=0.49\textwidth]{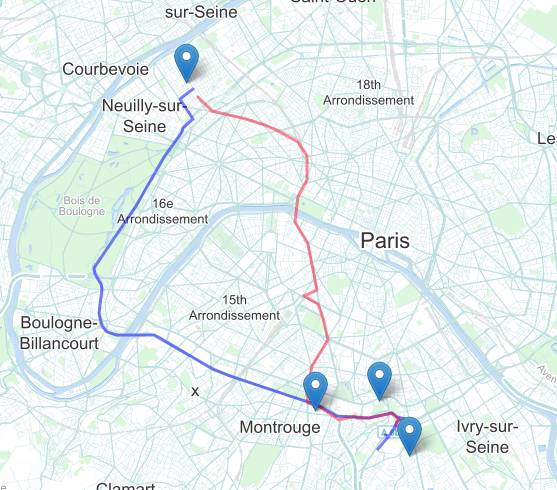}\label{fig:mf4}}
\caption{Illustration of some examples of mapping algorithm failure}
\end{figure*}

In this section some examples of \textit{CT-Mapper} failure are presented. In all these examples the blue markers are the cellular observations with the frequency of 15 minutes. The blue lines illustrate the real path and the red lines are the result of mapping algorithm. In Fig. \ref{fig:mf0} the real path and the result of mapping algorithm both are the on the road layer. 
The rest of three figures belong to the same daily commute of an individual in Paris. Fig. \ref{fig:mf2} is provided to show the example of correct matching, Fig. \ref{fig:mf3} is the case that a part of monomodal trajectory (on the road layer) is mapped over metro layer. Fig. \ref{fig:mf4} is another example of failure that the trajectory mistakenly mapped on the metro layer. From the failure examples, we can conclude that the result of our mapping algorithm can significantly improve if observations with higher frequency (e.g. 5 minutes instead of the 15 minutes) are provided.

\end{document}